\documentclass[twocolumn,eqsecnum,aps,pra,superscriptaddress,showpacs]{revtex4}

\usepackage{dcolumn}
\usepackage{graphicx}
\usepackage{epsf}
\usepackage{amsmath}
\usepackage{bm}
\usepackage{color}

\newcommand{\sss}{\scriptstyle}
\newcommand{\sssz}{\scriptstyle 0}

\newcommand{\dd}{{\mathrm d}}
\newcommand{\ii}{{\mathrm i}}
\newcommand{\ee}{{\mathrm e}}
\def\half{{\textstyle{\frac12}}}

\newcommand{\calC}{{\mathcal C}}
\newcommand{\calP}{{\mathcal P}}
\newcommand{\calT}{{\mathcal T}}

\newcommand{\q}{{\!\!\!}}

\newcommand{\p}{{\,\,\,\,\,\,}}
\newcommand{\m}{{\!\!\!\!\!\!}}

\newcommand{\C}{{p}}
\newcommand{\V}{\phantom{-}}

\begin{document}

\sloppy

\title{Quantum dynamics in atomic-fountain experiments for\\
measuring the electric dipole moment of the electron with improved sensitivity}

\author{B. J. Wundt}
\email{bjwcr7@mst.edu}
\affiliation{Department of Physics,
Missouri University of Science and Technology,
Rolla, Missouri 65409, USA}

\author{C. T. Munger}
\email{charlestmungerjr@gmail.com}
\affiliation{Stanford Linear Accelerator Center (SLAC),
Stanford, California 94309, USA}
\affiliation{Lawrence Berkeley National Laboratory (LBNL),
Berkeley, California 94720, USA}

\author{U. D. Jentschura}
\email{ulj@mst.com}
\affiliation{Department of Physics,
Missouri University of Science and Technology,
Rolla, Missouri 65409, USA}

\begin{abstract}
An improved measurement of the electron electric dipole moment (EDM) appears
feasible using ground-state alkali atoms in an atomic fountain in which a
strong electric field, which couples to a conceivable 
electron dipole moment (EDM),
is applied perpendicular to the fountain axis.  In a
practical fountain, the ratio of the atomic tensor Stark shift to the Zeeman
shift is a factor~$\mu\sim 100$.  We expand the complete time evolution
operator in inverse powers of this ratio;
complete results are presented for atoms of total spin~$F=3$, $4$,
and~$5$. For a specific set of entangled hyperfine sublevels (coherent states),
potential systematic errors enter only as even powers of $1/\mu$, making the
expansion rapidly convergent.
The remaining EDM mimicking effects are further suppressed in a 
proposed double-differential setup, where the final state 
is interrogated in a differential laser configuration,
and the direction of the strong electric field also is inverted.
Estimates of the signal available at existing accelerator facilities indicate that the
proposed apparatus offers the potential for a drastic improvement in EDM limits
over existing measurements, and for constraining the parameter space of
supersymmetric (SUSY) extensions of the Standard Model.\\
{\bf Subject Areas:}
Quantum Physics, Atomic and Molecular Physics, Particles and Fields
\end{abstract}


\maketitle

%
% INTRODUCTION AND OVERVIEW
%
\section{INTRODUCTION AND OVERVIEW}
\label{intro}

%
% Significance of the electron dipole moment
%
\subsection{Significance of the electron dipole moment}

A permanent electric dipole moment of the electron or of any other fundamental
particle, or of an atom in an eigenstate of angular momentum, is possible only
if the symmetries of both parity~($\calP$) and time reversal~($\calT$) are
violated \cite{PuRa1950,La1957}.  By the $\calC \calP \calT$~theorem
\cite{LuZu1958}, $\calT$ violation is the equivalent of $\calC
\calP$~violation, where~$\calC$ is charge conjugation.  $\calC \calP$~violation
has been observed, but only in the quark sector and only in the decay of the
neutral~$K$ and~$B$ mesons.  The Cabbibo-Kobayashi-Maskawa~(CKM) mass mixing
matrix in the Standard Model is consistent with these observations, but does
not produce a large enough $\calC \calP$ violating effect to account for the
excess of matter over anti-matter in the
universe~\cite{Sa1967,Sa1991,BeNiVo2004}.

Extensions of the Standard Model generically contain new massive particles and
new sources of $\calC \calP$~violation that give rise to an electric dipole
moment (EDM) for the electron. In field theories, the
interaction of an electron with the electromagnetic field is described
by the field-theoretical interaction Hamiltonian
$H_{\rm I}(t) = \int \dd^3 r \; {\mathcal H}(t, \vec r)$,
where the interaction Hamiltonian density
${\mathcal H} = {\mathcal H}(t, \vec r)$ is (in
international mksA (SI) units)
\begin{equation}
\label{toGamma}
{\mathcal H} = e \, c\,
\overline\psi \, \gamma^\mu \, \psi \, A_\mu \rightarrow e \, c\,
\overline\psi \,\Gamma^\mu (q) \, \psi \, A_\mu \,.
\end{equation}
Here $\gamma^\mu$ is a Dirac matrix,
$e = -|e|$ is the electron charge, $c$ is the speed of light,
and we assume that the fermion field operator $\psi$
is normalized so that $|\psi|^2$ has physical dimension
of inverse volume. The second form in Eq.~\eqref{toGamma} 
includes radiative
corrections, where the vertex function is denoted as $\Gamma^\mu$.
For an electron on the mass shell, the vertex function in an
arbitrary field theory can be expressed in terms of
Lorentz and $\calC \calP \calT$ invariant terms as
\begin{equation}
\label{F123}
\begin{split}
\Gamma^{\mu}(q) = {}&F_1 (q^2) \, \gamma^{\mu} +
F_2 (q^2) \frac{\ii}{2 m_{\ee} c} \sigma^{\mu \nu} q_{\nu} \\
{} +{}&  F_3 (q^2) \,
\frac{1}{2 m_{\ee} c} \, \sigma^{\mu \nu} \gamma^5 q_{\nu} \\
{}+{}& F_A (q^2) \,
\frac{G_F \,a}{8 \pi \sqrt{2}}
\left( \gamma^{\mu} \gamma^5 q^2 - 2 m_{\ee} c\, \gamma^5 q^{\mu} \right)  \,,
\end{split}
\end{equation}
where $\sigma^{\mu \nu} =
\ii \, [\gamma^\mu, \gamma^\nu]/2$, and $a$ is the contribution of
the anapole moment~\cite{DoKe1980}.
(The form factors are the $F_1$ Dirac, the $F_2$ Pauli,
and the $F_A$ anapole form factor; the $\calC \calP$-odd 
$F_3$ term which leads to the EDM does not carry a special name.)
The first and second terms in Eq.~\eqref{F123} 
conserve~$\calC$, $\calP$, and~$\calT$ separately.
The weak interaction gives rise to the last,
anapole term, which conserves~$\calT$ and~$\calC \calP$, but violates both~$\calP$
and~$\calC$ separately; it is not considered in the following.
The third term involves the form factor~$F_3$ and is odd
under~$\calP$ and~$\calT$, but conserves~$\calC$. In the non-relativistic limit,
the electron EDM gives rise
to the following effective Hamiltonian for the interaction of the
spin of a free electron with an electric field,
\begin{equation}
\label{defdipolemoment}
h_{\rm EDM} = (-d_{\ee}) \, \vec\sigma\cdot \vec E \,,
\qquad
d_{\ee} = -F_3(0) \, \frac{\hbar \, |e|}{2 m_{\ee} c} \,.
\end{equation}
In the effective Hamiltonian for a bound electron, one
has to replace $\vec\sigma \to R \, \vec F$ where 
$\vec F = (F_x, F_y, F_z)$ is the total angular momentum 
of the atom (including the nuclear spin) and $R$ is
the enhancement factor [see Eq.~\eqref{comp} below].
We recall that $|e| = -e$ is the elementary
charge unit.  The Pauli spin matrices $\vec \sigma$ are dimensionless.  The electric
field $\vec E$ is multiplied by the elementary charge and the electron Compton
wavelength $\hbar/(m_{\textrm e} c)$ to yield an interaction energy.  In the
absence of a special structure that would otherwise constrain the form
factor~$F_3$ to be small, field theories which extend the Standard Model
generally contain a significant EDM.  The special structure in the Standard
Model is that $\calC \calP$~violation occurs only in a single phase in the
quark mass mixing matrix. The leading Standard Model contribution to a lepton
EDM is at the level of four loops~\cite{PoRi2005}.  Because the resulting
electron~EDM is far too small to be observed by any proposed experiment, there
are in practice no Standard Model effects to account for, so mere observation
of an electron~EDM is direct evidence of a new and non-Standard-Model source of
$\calC \calP$~violation~\cite{WoTr2004}.

The sensitivity of EDM measurements to new phenomena is far-reaching.  In some
recent calculations, an electron~EDM arises through a mechanism that produces
the neutrino mass~\cite{MoEtAl2007} or is sensitive to physics at energy scales
that exceed $10^8\,{\rm GeV}$~\cite{PoRiSa2006}, or is sensitive to dark
matter~\cite{MaSe2006}, or that is responsible for baryogenesis~\cite{Se2006prd}.
The present limit on the electron~EDM already presses supersymmetry (SUSY),
especially when that limit is combined with those on the neutron~EDM and those
on the~EDMs of diamagnetic atoms. Present limits on the electron
EDM~\cite{HuEtAl2011,ReCoScDM2002} are lower by a factor of~$100$ than 
EDMs predicted by some
SUSY~models~\cite{FiPaTh1992,AbKhLe2001,Kh2003,OlPoRiSa2005,BeSu1991,Ba1993cp}
with super-partner masses of $100\,{\rm GeV}$ and $\calC \calP$ violating
phases of order unity.  Therefore, these SUSY models could be excluded.  The
present EDM limits are also not in complete agreement with models with one-TeV
super-partner masses~\cite{LiPrRa2010}.

In this paper, we examine in detail a proposal to significantly
lower experimental limits on the electron~EDM, by roughly two orders of
magnitude in comparison to present limits~\cite{HuEtAl2011,ReCoScDM2002}.  A
nonvanishing EDM on this level would imply the existence of new physics beyond
the Standard Model, or, alternatively, imply that currently favored extensions
of the standard model need to be significantly altered. Within~SUSY, an
unexpectedly  small EDM could be realized if one assumes that $\calC \calP$
violating phases are much smaller than currently assumed, or that the masses of
superpartners are much larger than currently expected. Quite sophisticated
models have been proposed with this notion in mind: E.g., in so-called split
Supersymmetry~\cite{ArDiGiRo2005,ChChKe2005,GiRo2006}, one assumes that only
the masses of those superpartners that contribute to EDMs most significantly,
are larger than expected, whereas the masses of other SUSY partners remain in
the expected range. However, in general, SUSY offers no special reason for
small $\calC \calP$~violating phases or any good reason for most of the
superpartner masses to remain small~\cite{Ha2008}.  If more accurate
experimental results still turn out to be compatible with a zero EDM, one will
begin to exhaust some of the simpler remedies and push the theory towards
constructions inconsistent with the original motivations for~SUSY.

%
% Experimental idea and overview
%
\subsection{Experimental idea and overview}

Our proposed measurement scheme is based on the observation that the
interaction Hamiltonian~\eqref{defdipolemoment} is proportional to $\vec\sigma
\cdot \vec E$ [or $\vec F\cdot \vec E$ for a bound electron, see
Eq.~\eqref{comp} below].
The influence of an EDM of the electron is largest
when an atom is put in an intense, static electric field to evolve a long time,
and an atomic fountain apparatus (the details of which will be explained below) 
is an essentially undisturbed environment in
which this can be done. The inclusion of an atomic fountain
in EDM experiments has not been discussed in the literature to
the best of our knowledge.
However, it is a subtle matter to select an actual
observable that is sensitive to an EDM, takes advantage of the properties of an
atomic fountain, cancels many systematic effects, and can be realized using
lasers alone.  Our proposed scheme consists of four steps.

{\bf [Step 1.]} We start with heavy alkali atoms
of half-integer nuclear spin
and prepare an initial state that is a coherent superposition
of states within the upper hyperfine manifold of the
electronic $S_{1/2}$ ground state.
 
{\bf [Step 2.]} In the fountain, the state of the atom evolves from the initial one 
under the influence of a strong electric
field but only weak static magnetic fields.
The full Hamiltonian generating the dynamics is 
given below in Eq.~\eqref{Htotal}.
The quantum dynamics of the atom are the main
concern of the current theoretical investigation.
Because of the analytic simple structure of the Hamiltonian 
given in Eq.~\eqref{Htotal}, which involves a tensor Stark term 
and an EDM term, the dynamics 
can be described semi-analytically using time-ordered
perturbation theory.

{\bf [Step 3.]} The analysis of the time-evolved state proceeds in a region where the
atom is irradiated by laser light tuned to the $S_{1/2} \big| F,M \big> \to
P_{j'}\big|F'=j'+1/2, M'\big>$ transition 
(with $j' = 1/2$ or $3/2$ and $F = I + 1/2$), 
roughly as follows (further details are discussed in the main 
body of the article):
After many cycles of absorption and spontaneous emission,
any sublevel $S_{1/2}\big| F,M \big>$ either ends in the dark state of the
upper hyperfine level ($F=I+1/2$) with probability~$p_M$ or ends in the lower
hyperfine level ($F=I-1/2$) with probability~$1-p_M$. 
If an atom ends in the dark state, the atom ``survives'' in the
upper hyperfine level. The probability that
the actual time-evolved state survives
is given as the sum over the ``survival probabilities'' of its separate components.
This probability depends on the electron EDM and defines the 
observable
$S(\theta)$ [see Eq.~\eqref{defRz} below], where $\theta$ is the 
rotation angle (about the $z$ axis in Fig.~\ref{figg1} below)
of the propagation direction of the linearly polarized analysis laser with regard
to the coordinate system of the atomic fountain apparatus
(see Fig.~\ref{figg1} below for the $x$, $y$ and $z$ axes).
We here assume that the atomic state is initially 
prepared with respect to a quantization axis adjusted to be essentially parallel 
to the direction of the applied electric field
(which is the $z$ axis in Fig.~\ref{figg1} below).
 
{\bf [Step 4.]} As a separate step, the
survival probability signal $S(\theta)$ is measured
by fluorescing the $S_{1/2},F=I+1/2$ to $P_{3/2},F'=I+3/2$ cycling transition,
then pumping all atoms from the lower $S_{1/2},F=I-1/2$ hyperfine level also into the
$S_{1/2},F=I+1/2$ level and fluorescing the cycling transition again.  The ratio
of the number of photons scattered over a fixed time while the cycling transition
was fluoresced then equals the desired probability.
Throughout this paper, 
we use the phrase ``optical pumping" when the purpose of hitting
atoms with a laser is to alter the atomic state, and "fluorescing" when the
purpose is to scatter photons from the laser that are to be counted.

We propose to do all of this in a double-differential
setup, where the interrogation laser in step~$3$
is tilted by angles~$\theta$ and~$-\theta$, and the difference 
of the two survival probabilities
$P(\theta) = S(\theta) - S(-\theta)$ is taken
[see also Eq.~\eqref{defP} below]. 
That difference is measured for opposite signs of the applied
electric field and the difference is taken again.
This leads to the observable $P^o(\theta)$
(the superscript standing for ``odd'') that appears in
Eq.~\eqref{defPo} below.
This double-differential setup is key to our proposal
and eliminates many systematic effects.

A schematic diagram for the resulting experiment to measure the electron
electric dipole moment using an atomic fountain is laid out in
Fig.~\ref{figg1}. Atoms are collected and cooled in a magneto-optical trap, a
fully coherent initial state is prepared
(so that the use of a density matrix becomes unnecessary), and the atoms are launched
vertically.  The atoms enter a region shielded against static magnetic fields.
Within the atomic fountain, the atoms then rise into and fall out of a large
electric field. If an electron EDM exists, the initial quantum state (step~$1$)
rotates by a small angle about the electric field axis, while in the electric
field (step~$2$).  While still in the magnetically shielded region, but outside
the electric field, the atoms are analyzed by optical pumping (state analysis,
step~$3$), which has the effect of encoding the small angle of rotation into a
relative shift in the population of the upper and lower hyperfine levels of the
ground state.  Those populations don't change if stray magnetic fields are
subsequently applied, so atoms can be allowed to fall out of the magnetic
shield before the populations are measured by optical pumping and counting of
the scattered photons (fluorescence, step~$4$).

\begin{figure}[t!]
\begin{center}
\includegraphics[width=0.93\linewidth]{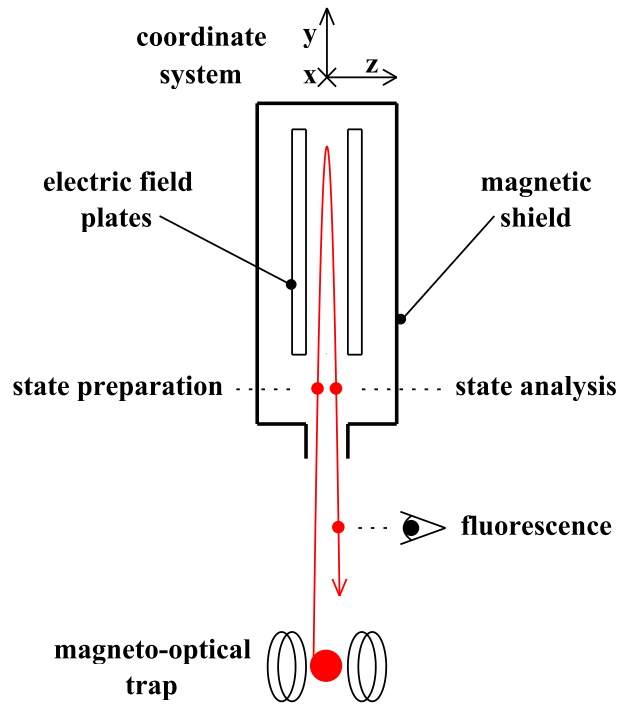}
\end{center}
\caption{\label{figg1} Schematic diagram of an experiment to measure an
electron EDM using an atomic fountain.  In the diagram the width of the
parabolic trajectory has been greatly exaggerated. In first approximation,
we assume that
atoms rise and fall strictly along the $y$~axis, and that the applied electric
field is strictly parallel to the $z$~axis.  Atoms are exposed to lasers at the
points on the trajectory marked by solid circles.}
\end{figure}

The system of optical pumping for state preparation uses the property that an
atom illuminated with a laser tuned to an allowed transition with $\Delta F =
0$ has a unique, known ``dark'' state from which excitation by the laser is
forbidden.  The dark state is a continuous function of the ellipticity of the
laser polarization; for a linearly polarized laser with its axis of
polarization parallel to the~$z$~axis, the dark states is  $\big|F,0\big>$, and
for a laser propagating in the $z$~direction that is circularly polarized with
positive helicity, the dark state is $\big|F, F\big>$.  For conceptual
simplicity, we mainly study initial states which are the coherent
superpositions $(\big|F,M\big>\pm \big|F,-M\big>)/\sqrt{2}$.

\begin{figure}[t!]
\begin{center}
\includegraphics[width=0.93\linewidth]{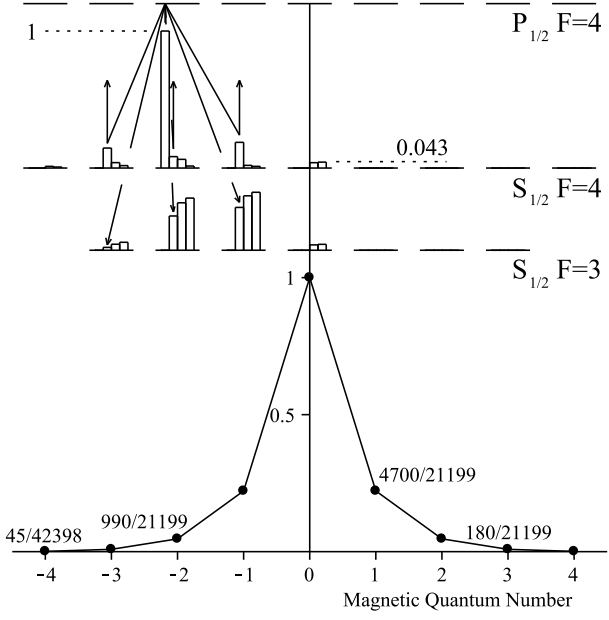}
\end{center}
\caption{\label{figg2}
An atom, initially in the state $S_{1/2}\big|4, M\big>$,
undergoes a large number of cycles of absorption of laser photons,
followed by spontaneous emission.
We define the probability~$\C_M$ to be the probability
that an atom, illuminated with a $\pi$-polarized
laser tuned to the transition $S_{1/2},\,F=4$ to $P_{1/2},\,F=4$,
ends up in the dark $S_{1/2}\big|4,0\big>$ state instead of
ending in some state in the $S_{1/2},\,F=3$ hyperfine level,
after many cycles of absorption and spontaneous emission.
In the upper half
of the figure, the arrows show the pattern for the first spontaneous emission
and re-excitation for an atom initially in state $S_{1/2}\big|4,-2\big>$, and
the histograms show how the probability of being in any given level evolves
after $0$, $1$, $2$, or $3$ spontaneous emissions.
The probability of ending in the dark  $S_{1/2}\big|4,0\big>$
state involves processes like
$S_{1/2}\big|4,-2\big> \to P_{1/2}\big|4,-2\big> \to
S_{1/2}\big|4,-1\big> \to P_{1/2}\big|4,-1\big> \to
S_{1/2}\big|4,0\big>$ and similar.
Values for $p_M$ are in the graph in the lower half of the figure,
and correspond to those given in the second column of 
Table~\ref{table1}.}
\end{figure}

The system of optical pumping for state analysis
(step~$3$) is similar to that of state
preparation (step~$1$).
An atom in any state of the upper, $F=I+1/2$ hyperfine level that
is illuminated with a laser tuned to an allowed transition with $\Delta F =0$
will either end in the dark state of the laser or in some state of the lower
hyperfine level.  In the simple cases of a laser that is polarized parallel to
the~$z$~axis, or that propagates parallel to~$z$ and is circularly polarized,
we define the probability~$\C_M$ that an atom originally in the state
$\big|F, M\big>$ ends in the dark state
after multiple cycles of excitation to the $P_{j'}$ state
and spontaneous emission.
For $\pi$~polarization, the dark state is $\big| F,0\big>$,
because $(M=0)$--($M=0$) transitions are only allowed if the
total angular momentum changes.
The probabilities~$\C_M$ depend on the
branching ratios for the various spontaneous transitions and can be found as
simple fractions by solving the $2F+1$ equations for the evolution of the
states of a given hyperfine level; the essential physics for the probabilities
is illustrated for one particular transition in Fig.~\ref{figg2}.
In the calculation, the probability that an atom of nuclear spin $I$
in a $P$-state with quantum numbers $j',F',M'$
undergoes a radiative transition to an $S_{1/2}$ state
with quantum numbers $j,F,M$ is proportional to
\begin{equation}
(2j' +1)^2 \, (2F+1) \\[2ex]
\left( \begin{array}{ccc} j & 1 & j' \\
m & m'-m & -m' \end{array} \right) \,
\left\{ \begin{array}{ccc} F & 1 & F' \\
j' & I & j \end{array} \right\} \,,
\end{equation}
where standard notation is used for the $3j$ and $6j$ symbols.
One then sets up and solves the rate equations.
In the rate equations, one may ignore the small frequency
difference between the upper and lower hyperfine manifolds
of the ground state $S_{1/2}\big|I\pm 1/2,M\big>$
indicated in Fig.~\ref{figg2}, because the frequency
difference to the upper state $P_{j'}$ is much larger
than the hyperfine splitting. The sets of
values $\C$ that are achievable for the case $F=4$ are listed in
Tables~\ref{table1} and~\ref{table2};
these turn out to be rational numbers.
Values for $F=3$ and $F=5$
are found in Appendix~B of Ref.~\cite{WuMuJe2012app}.

\begin{table}[t]
\caption{\label{table1}
For $I=7/2$, the probability~$p_M$ that an atom, initially in
the state $S_{1/2}\big|4, M\big>$, remains in the $F=4$ hyperfine level after
being pumped with light tuned to the transition $S_{1/2},\,F=4\to P_{j'},\,F'$
that is linearly polarized with the axis of polarization parallel to~$z$.}
\begin{ruledtabular}
\begin{tabular}{ccccc}
$P_{j'}$ & $P_{1/2}$ & $P_{3/2}$ & $P_{1/2}$ & $P_{3/2}$\\
$F'$ & $4$ & $4$ & $3$ & $3$\\[0.5ex]
$S_{1/2}\big|4, M\big>$ & & & & \\
\colrule
$\vphantom{\big|_{\big|}^{1}}$$M=4$ & $\frac{45}{42398}$ &
$\frac{151263}{23049298}$ & $1$ & $1$\\
$\vphantom{\big|_{\big|}^{1}}$$M=3$ & $\frac{180}{21199}$ &
$\frac{345744}{11524649}$ & $\frac{1792}{2585}$ & $\frac{839216}{4101383}$\\
$\vphantom{\big|_{\big|}^{1}}$$M=2$ & $\frac{990}{21199}$ &
$\frac{1205694}{11524649}$ & $\frac{1092}{2585}$ & $\frac{133868}{4101383}$\\
$\vphantom{\big|_{\big|}^{1}}$$M=1$ & $\frac{4700}{21199}$ &
$\frac{3788344}{11524649}$ & $\frac{112}{517}$ & $\frac{15680}{4101383}$\\
$\vphantom{\big|_{\big|}^{1}}$$M=0$ & $1$ & $1$ &
$\frac{70}{517}$ & $\frac{2450}{4101383}$\\
$\vphantom{\big|_{\big|}^{1}}$$M=-1$ & $\frac{4700}{21199}$ &
$\frac{3788344}{11524649}$ & $\frac{112}{517}$ & $\frac{15680}{4101383}$\\
$\vphantom{\big|_{\big|}^{1}}$$M=-2$ & $\frac{990}{21199}$ &
$\frac{1205694}{11524649}$ & $\frac{1092}{2585}$ & $\frac{133868}{4101383}$\\
$\vphantom{\big|_{\big|}^{1}}$$M=-3$ & $\frac{180}{21199}$ &
$\frac{345744}{11524649}$ & $\frac{1792}{2585}$ & $\frac{839216}{4101383}$\\
$\vphantom{\big|_{\big|}^{1}}$$M=-4$ & $\frac{45}{42398}$ &
$\frac{151263}{23049298}$ & $1$ & $1$\\
\end{tabular}
\end{ruledtabular}
\end{table}

\begin{table}[t]
\caption{\label{table2}
For $I=7/2$
the probability~$p_M$ that an atom
initially in the state $S_{1/2}\big|4, 4\big>$ remains
in the $F=4$ hyperfine level after being pumped with light tuned to the
transition $S_{1/2},\,F=4\to P_{j'},\,F'$, that propagates in the $z$
direction, and is circularly polarized with positive helicity.}
\begin{ruledtabular}
\begin{tabular}{ccccc}
$P_{j'}$ & $P_{1/2}$ & $P_{3/2}$ & $P_{1/2}$ & $P_{3/2}$\\
$F'$ & $4$ & $4$ & $3$ & $3$\\[0.5ex]
$S_{1/2}\big|4, M\big>$ & & & & \\
\colrule
$\vphantom{\big|_{\big|}^{1}}$$M=4$ & $1$ & $1$ & $1$ & $1$\\
$\vphantom{\big|_{\big|}^{1}}$$M=3$ & $\frac{4}{11}$ &
$\frac{28}{53}$ & $1$ & $1$\\
$\vphantom{\big|_{\big|}^{1}}$$M=2$ & $\frac{80}{451}$ &
$\frac{3248}{10123}$ & $\frac{35}{47}$ & $\frac{35}{143}$\\
$\vphantom{\big|_{\big|}^{1}}$$M=1$ & $\frac{1468}{17589}$ &
$\frac{352996}{1791771}$ & $\frac{469}{705}$ & $\frac{1141}{6721}$\\
$\vphantom{\big|_{\big|}^{1}}$$M=0$ & $\frac{14774}{334191}$ &
$\frac{3868942}{30460107}$ & $\frac{71}{141}$ & $\frac{6965}{154583}$\\
$\vphantom{\big|_{\big|}^{1}}$$M=-1$ & $\frac{7340}{334191}$ &
$\frac{2470972}{30460107}$ & $\frac{1037}{2679}$ & $\frac{186935}{10357061}$\\
$\vphantom{\big|_{\big|}^{1}}$$M=-2$ & $\frac{155080}{13033449}$ &
$\frac{288122744}{5391438939}$ & $\frac{7883}{29469}$ & $\frac{1867985}{445353623}$\\
$\vphantom{\big|_{\big|}^{1}}$$M=-3$ & $\frac{3196660}{534371409}$ &
$\frac{35621245604}{1029764837349}$ & $\frac{4771}{29469}$ &
$\frac{15510145}{18259498543}$\\
$\vphantom{\big|_{\big|}^{1}}$$M=-4$ & $\frac{18319475}{5878085499}$ &
$\frac{72657047561}{3210443316441}$ & $\frac{688}{9823}$ &
$\frac{46289600}{529525457747}$\\
\end{tabular}
\end{ruledtabular}
\end{table}

For the cases of linear and circular polarization, the probability that an atom,
emerging from the atomic fountain
in a general superposition of states $\sum A_M \big|F, M\big>$ for
amplitudes~$A_M$, ends in the dark state is simply $\sum |A_M|^2 \, \C_M$.  Such a
probability encodes information about the original atomic state in question and
can be easily measured using fluorescent detection.
Final read-out and normalization to the
incoming flux of atoms proceeds as follows
(step~$4$): A cycling transition
from the $S_{1/2}$ ground state with spin $F=I+1/2$ to the $P_{3/2}$~state with $F=I+3/2$
is rung and the photons counted; then a laser tuned to the transition from the
lower hyperfine level of the
ground state with $F=I-1/2$
to a $P$~state with $F=I+1/2$ puts all
atoms into the $S$-state
hyperfine level with $F=I+1/2$, and then the cycling transition
$S_{1/2} \to P_{3/2}$
is fluoresced and the photons counted again; the ratio of the counts gives the
probability.

An alkali atom of large nuclear charge would be
indicated for our atomic fountain because
relativistic effects enhance the~EDM of such an atom compared to that of the
free-electron by a large multiplicative factor~$R$ (the ``enhancement
factor'').  The theory of the enhancement factor is well-established, and has
been used to set the EDM limits in Ref.~\cite{PDG2010}.  Computed values of the
enhancement factors for Rb, Cs, and Fr are presented in Table~\ref{table3}.
None of these computed factors have varied by more than $20\%$ from the
earliest factors computed in~1966 (Ref.~\cite{Sa1966}),
and an experiment to discover an EDM does
not depend upon the error in the computed enhancement factor being small.
Estimates of the size of limit that may be set by a francium fountain
experiment operated at existing accelerator facilities may be found in
Appendix~A of Ref.~\cite{WuMuJe2012app}.

\begin{table}[b]
\caption{\label{table3}%
Calculations of the enhancement factor
$R$ for the ground state of heavy alkali atoms.}
\begin{ruledtabular}
\begin{tabular}{cdcc}
\textrm{Alkali}&
\textrm{$R$}&
\textrm{Year}&
\textrm{Reference}\\
\colrule
    &24               & 1966 & \cite{Sa1966} \\
Rb  &24.6             & 1985 & \cite{JoGuIdSa1986} \\
    &25.68            & 1994 & \cite{ShDaAn1994}   \\
    &25.74            & 2008 & \cite{NaSaDaMu2008}  \\
\colrule
    &119              & 1966 & \cite{Sa1966}  \\
    &114.9            & 1985 & \cite{JoGuIdSa1986, JoGuIdSa1985} \\
Cs  &114              & 1990 & \cite{MPOe1987, HaLiMP1990} \\
    &120.54           & 2008 & \cite{NaSaDaMu2008} \\
    &124              & 2009 & \cite{DzFl2009}\\
\colrule
&1150.\footnote{This early value did not include a correction for the shielding
factor of the atomic core.  The addition of a shielding correction lowers the
enhancement factor of all other alkali atoms.}
    & 1966 & \cite{Sa1966} \\
Fr  &910              & 1999 & \cite{Byrnes1999} \\
    &894.93           & 2009 & \cite{Mukherjee2009} \\
\end{tabular}
\end{ruledtabular}
\end{table}

A key part of any EDM measurement is controlling any systematic error that,
like the effect of the EDM, reverses sign when the electric field is reversed.
Such effects arise naturally in EDM experiments because in the rest frame of
the atom, due to the atom's nonzero velocity in the applied electric field, the
atom sees (SI units) a magnetic field
\begin{equation}
\vec B_{\rm mot} = \frac{1}{c^2}\, \vec v\times \vec E \,,
\end{equation}
which changes sign when $\vec E$ changes sign.
The interaction of the atom's magnetic moment with that motional
magnetic field (Zeeman effect) then also changes sign
when the direction of $\vec E$ is reversed.
The rotation of the atom by this motional magnetic field,
necessarily perpendicular to the electric field, cannot by
itself generate a rotation about the electric-field axis.
{\em A priori}, one would thus assume that the
motional magnetic field does not mimic an EDM.
However, in any practical apparatus, there will inevitably
be present trace static magnetic fields which the atom will explore
as it moves; the combinations of rotations about
these fields and about the motional field are of concern.

In this paper, we present a complete categorization of all systematic errors
that result from an atom in an atomic fountain being subjected to a constant
electric field, the motional magnetic field, and trace magnetic fields which
are static but may vary arbitrarily in all three spatial directions.  The
unitary operator that gives the time evolution of any superposition of magnetic
sublevels~$\big|F,M\big>$ is a complex matrix of
dimension~$(2F+1)\times (2F+1)$, where~$F$ is
the total spin of the hyperfine level.  In an atomic fountain, the maximum
atomic velocity~$\sim 4\,{\rm m/s}$ is two orders of magnitude smaller than in
previous atomic beam experiments~\cite{ReCoScDM2002}, so the motional magnetic
field also is reduced by two orders of magnitude. In an atomic fountain,
atoms pass through each point in space twice, once rising and once falling; the
resulting atom-by-atom reversal of the atomic velocity cancels some
motional-field dependent systematics without a need for a second,
antipropagating atomic beam. 
Let us briefly comment on the effect of a possible slight
nonuniformity of the electric field, in which case the atomic trajectory 
may become more complicated because
atoms in the level $S_{1/2}, F=I+1/2$ are attracted into regions of strong electric field by the Stark shift. 
However, in our differential setup, we plan to measure the 
signal twice, with the sign of the electric field inverted, and in this 
case, even a more complicated trajectory is conserved and independent of the state of the atom, 
in the approximation that $\alpha_F^S \gg \alpha_F^T$ in Eq.~\eqref{polarizability}.

Moreover, under the conditions of an atomic
fountain, the shift of the magnetic sublevel~$\big|F,M\big>$ due to the tensor
Stark effect is now much greater than its shift due to the Zeeman effect.  In a
constant electric field, the unitary operator that gives the time evolution of
any state can be computed as a series in reciprocal powers of a dimensionless
parameter~$\mu\sim 100$, which represents (roughly) the ratio of the shifts.
The $(1/\mu)$-series therefore is rapidly convergent.  We present this series
as the sum of analytic integrals over the magnetic fields, times matrices of
constants, for any of the total spins~$F = 3$, $4$, and~$5$, and so for the
spins of all the alkali atoms of experimental interest.  We place special
emphasis on the case $F=4$.  A key result is that a properly constructed
observable sensitive to an EDM will contain errors that are only {\sl even}
powers of~$1/\mu$. Because contributions of order~$1/\mu^4$ prove negligible in
practice, any experiment only has to control the few terms of
order~$1/\mu^2$.

The coordinate system used to describe the atomic fountain is shown in
Fig.~\ref{figg1}.  The atom remains on a single vertical
axis, the~$y$~axis. The direction of the electric field is horizontal and
defines the~$z$~axis. Therefore, the motional magnetic field
is parallel to the~$x$~axis.  The effective Hamiltonian within a
manifold of states of total spin~$F$ is the sum of several pieces.  The
contribution of the electron EDM is
\begin{equation}
\label{comp}
H_{\rm EDM} = R\, ( -d_{\ee}) \, \vec F \cdot \vec E \,,
\end{equation}
where~$\vec F$ is the total angular momentum of the atom,
$d_{\ee}$ is the electron EDM, and~$R$ is the
enhancement factor.  The contribution of the Zeeman effect is
\begin{equation}
H_{\rm Z} = \mu_B \, g_F \, \vec F\cdot \vec B \,,
\end{equation}
where~$\mu_B = \hbar |e|/(2 m_e c)$ is the Bohr magneton and~$g_F$ is the Land\'e
$g$ factor for the manifold.  The Stark effect contributes
to each level~$\big|F,M\big>$ the energy shift
\begin{equation}
{\cal E}_{\rm S} = -\frac{1}{2}(\alpha_0 + \alpha_{FM}) \, E_z^2\,,
\end{equation}
where $\alpha_0$ is the scalar polarizability (which is independent
of~$F$ and~$M$) and~$\alpha_{FM}$ is the tensor polarizability.
The tensor polarizability splits~\cite{DzEtAl2010,Sa1967Stark} into the sum
\begin{equation}
\label{polarizability}
\alpha_{FM} = \alpha^S_F + \frac{3 M^2-F(F+1)}{F(2F-1)}\alpha^T_F \,,
\end{equation}
where all dependence on the magnetic quantum number~$M$ is explicit.  The parts
of~$\alpha_0$ and of~$\alpha_{FM}$ that are independent of~$F_z$ contribute to
a global shift of the whole hyperfine manifold and thus introduce no change in
the atomic state other than a global phase.  We may therefore drop them in
solving for the time evolution of the atomic state. In doing so, we are well
aware of the fact that, while the shifts do not affect the state, they still
have a large effect on the motion of the atom. In general, the global
shift of the hyperfine manifold implies that an atom
accelerates as it enters an electric field.  Furthermore, a parallel beam of
atoms defocuses as it enters the electric field,
because atoms are pulled towards
the high-field region at the edges of the plates.  Such defocusing can be
compensated by a suitable set of electrostatic
lenses~\cite{KaLaGo2002,KaAmGo2005,Ka2011}.

Isolating the terms that are relevant for the quantum
dynamics (mixing within the~$F$-manifold), this gives an effective Hamiltonian
\begin{equation}
H_{\rm S} = A_{\rm S} \, E_z^2 \, F_z^2 \, ,
\end{equation}
where $A_{\rm S}$ is the constant
\begin{equation}
A_{\rm S} = -\frac{3 \, \alpha_F^T}{2F(2F-1)} \, .
\end{equation}
The total effective Hamiltonian $H_{\rm t}$ then is the sum of the
tensor Stark term, the EDM term, and the Zeeman term,
\begin{equation}
\label{Htotal}
H_{\rm t} = A_{\rm S} \, F_z^2 \, E_z^2+ \mu_B \, g_F\,  \vec F \cdot \vec B
- R \, d_{\rm e} \, F_z \, E_z \, .
\end{equation}
Contributions to the effective Hamiltonian
from the mixing of different hyperfine levels or from terms
in the Stark effect of order $E_z^4$
(the hyperpolarizability~\cite{FuRe1993}) are negligible.
A remark about the units in the Hamiltonian is necessary here.
We have deliberately scaled the quantities in the
Hamiltonians so that the total
angular momentum $\vec F$ as well as the projection
quantum number $F_z$ of a quantum state are dimensionless.
To this end, we have absorbed $\hbar$ as the unit of the
angular momentum into the respective moments
and into the constant~$A_{\rm S}$,
using Eq.~\eqref{defdipolemoment} for the EDM
and using the usual definition of~$\mu_B$.

The atoms enter the electric field and spend a time~$T$ within it. In this
paper, we do not consider effects due to a continuous rise of the electric
field from zero, but model the rise as a step-function.  In principle, the
finite size of the transition region over which the electric field rises from
zero to its full value will contribute to the time evolution operator in a
rapidly convergent series in powers of the small time atoms spend in the
transition region. The resulting systematics can be controlled by reducing
static magnetic fields only in the transition region, whose vertical length
will only be a few centimeters, without having to reduce the static magnetic
fields everywhere in the whole fountain, whose vertical length will be about a
meter. In order to simplify the calculations,
we also introduce a dimensionless variable~$t$ that
runs from~$-\tau$ to~$\tau$, with~$\tau = 1/2$, while the usual time runs
from zero to~$T$.  We then define a standard electric field strength~$E_S$
by
\begin{equation}
E_S = \sqrt{\frac{\hbar}{T \, A_{\rm S} }} \,,
\end{equation}
in order to render the Schr\"odinger equation dimensionless. 
(Note: to use this equation the units
of $A_S$ have to be $J/(V/m)^2$.) The atom now evolves
according to $H(t)\psi={\rm i}\,
\partial\psi/\partial t$, where the Hamiltonian $H(t)$
is obtained from $H_{\rm t}$ given in Eq.~\eqref{Htotal}
by an appropriate scaling,
\begin{equation}
\label{Htt}
H(t) = \epsilon_z^2(t) \, F_z^2 +
\vec\beta(t)\cdot \vec F + \sigma_F \, \epsilon_z(t) \, F_z \,.
\end{equation}
Now $H(t)$ is dimensionless, and the (dimensionless) coefficients are
\begin{equation}
\label{sigmaF}
\epsilon_z(t) = \frac{E_z(t)}{E_S} \,, \ \,\,\,\, \vec{\beta}(t) =
\frac{g_F \, \mu_B \, \vec{B}(t)}{A_{\rm S} \, E_S^2} \,, \ \,\,\,\,
\sigma_F = - \frac{d_{\textrm e} \, R}{F \, A_{\rm S} \, E_S} \,.
\end{equation}
The time-dependent Hamiltonian~\eqref{Htt} is the basis of the entire
derivation that follows.  When the electric field is a
constant in time, we define a time-independent parameter
\begin{equation}
\label{defmu}
\mu=\epsilon_z^2 \,,
\end{equation}
which for a typical experimental setup has values in the range
$\mu\sim 100$.

%
% Time--Ordered Perturbation Theory
%
\section{Time--Ordered Perturbation Theory}
\label{modelcesium}

%
% Hamiltonian and time evolution
%
\subsection{Hamiltonian and time evolution}

Specializing the above general statements to an atom with a defined quantum
number~$F$, we now study the effective Hamiltonian in Eq.~\eqref{Htt} for
cesium (${}^{133}$Cs), whose upper hyperfine level has $F=4$.  As a simple
model, consider the fountain of Fig.~\ref{figg1} with the atoms confined to the
$y$ axis, with an electric field which might vary in magnitude but is
always parallel to the $z$ axis, and a magnetic field which in the atom's rest
frame varies only slowly in time. We define $t$ as a dimensionless time
variable. The ordinary (dimensional) time~$\tilde{t}$, as measured by a clock,
is related to $t$ as follows,
\begin{equation}
t = \left( \frac{\tilde{t}}{T} -\frac12 \right) \,.
\end{equation}
The turning point of the atoms is at $\tilde t = \tfrac12 \, T$.
The atoms enter the fountain at  $\tilde t = 0$, which corresponds to
the scaled time variable
$t=-\half$, and they leave the fountain at $t=+\half$. In the following, we employ the
more general notation
\begin{equation}
-\tau \leq t \leq \tau \,,
\qquad
\qquad
\tau=\half \,.
\end{equation}
This use of the symbol $\tau$ allows us to recognize
time-symmetries in the calculation. For example,
the motional magnetic field in the rest frame of an atom is odd around $t=0$ because
the velocity changes sign there, while
laboratory static magnetic fields are even in~$t$.

The Hamiltonian~\eqref{Htt} gives rise to a time evolution operator.  Complete
knowledge of this operator would allow us to propagate an initial state through
the fountain and obtain the final state, which determines the observables.  All
effects which are either larger or about the size of the target sensitivity
of~$2\cdot 10^{-50}\,{\rm C\,m}$ have to be
calculated if they could possibly  mimic an electron EDM.  At this stage, it is
useful to recall that the traditional unit for the electron EDM is $e\,{\rm
cm}$; the conversion to SI units is $1\,{\rm C\,m} = 1.602\cdot
10^{-21}\,e\,{\rm cm}$.  At the target sensitivity, the corresponding
dimensionless parameter $\sigma_F$ in our Hamiltonian \eqref{Htt} has the value
$\sigma_F = 4 \times 10^{-9}$ for cesium.

In the model calculation we consequently need to find all effects
that would lead to an EDM-like signal of a magnitude greater
than that of $\sigma_F$. Without the EDM term, the Hamiltonian reads as
\begin{equation}
\label{H0}
H_0(t) = \epsilon_{z}^2 (t) \, F_{z}^2 + \vec{\beta}(t) \cdot \vec{F} \,.
\end{equation}
The time-evolution governed by this Hamiltonian is characterized by the equation
\begin{equation}\label{eomU}
H_0(t) \, U_0(t) = \ii \, \frac{\partial}{\partial t} U_0(t) \,,
\end{equation}
with the initial condition $U(- \tau) = 1 $.

We first split this time evolution operator~$U(t)$
into a diagonal part~$V(t)$ and a remainder term~$W(t)$. For~$W(t)$, we define an
equation of motion and write it as a time-ordered exponential
$W(t) = {\bm T} \exp(-\ii \int_{-\tau}^{t} h(t') \dd t')$,
where $h(t)$ is defined in Eq.~\eqref{Hprimedef} below,
and ${\bm T}$ denotes the time ordering (recall the symbol $T$ denotes instead the
total time the atom spends in the fountain). Because $W(t)$
is expressed in terms of highly oscillatory integrals, for a constant magnetic field
it is amenable to
an expansion in the parameter~$1/\mu$ where $\mu$ is defined in Eq.~\eqref{defmu},
with coefficients that depend on time-integrals over various components of the
magnetic fields.

In the absence of magnetic fields in the $x$ and $y$ direction, the
Hamiltonian $H_0(t)$ becomes
\begin{equation}\label{H0forV}
H'_0(t) = \epsilon_{z}^2 (t) \, F_{z}^2 + \beta_z(t) \, F_z \,.
\end{equation}
This Hamiltonian is diagonal in the hyperfine manifold,
and its time-ordered exponential
\begin{equation}\label{Veom}
V(t) = {\bm T} \exp\left( -\ii \int_{-\tau}^{t} H'_0(t') \dd t' \right)
\end{equation}
is a diagonal $(2F+1) \times (2F+1)$ matrix
acting on a $2 F+1$ dimensional subspace of~$M$~levels
inside the manifold of a given total spin~$F$, which can be written as
\begin{widetext}
\begin{equation}
\label{Vt}
V(t) = \mathrm{diag} \!
\left(\!\!\!\!\begin{matrix}
\,\,\,\exp\Bigl[{ - \ii\! \int\limits_{-\tau}^{t} \!\left( F \beta_z(t') + 
F^2 \epsilon_z^2(t') \right) \dd t' }\Bigr]\qquad\qquad\; \\[2ex]
\,\exp\Bigl[{- \ii \! \int\limits_{-\tau}^{t} \!
\!\left( (F\!-\!1) \beta_z(t')\!+\! (F\!-\!1)^2 \epsilon_z^2(t') \right) \dd t' } \!\Bigr] 
\\[1ex] \vdots \\[1ex]
\!\!\exp\Bigl[{ - \ii\!\! \int\limits_{-\tau}^{t}
\!\left( (-F) \beta_z(t') + (-F)^2 \epsilon_z^2(t') \right) \dd t' }\Bigr] 
\end{matrix}\!\!\!\right) \,,
\end{equation}
where $\mathrm{diag}$ denotes the diagonal matrix generated by the
given matrix elements.
It is now our aim to solve the full problem by writing the full time-evolution
operator~$U_0(t)$ as the product of the time-evolution operator for the
diagonal part~$V(t)$ and the remainder term~$W(t)$, {\em i.e.}
\begin{equation}
U_0(t) = V(t) \, W(t) \,.
\end{equation}
The required equation of motion for $W(t)$ can be found by using this product
for $U(t)$ in Eq.~\eqref{eomU}, which gives
\begin{equation}
H_0(t) \, V(t) W(t) = \ii \,
\frac{\partial}{\partial t} \left( V(t) W(t) \right) \,, 
\qquad
H_0(t) V(t) W(t) - \ii \frac{\partial V(t)}{\partial t} W(t) =
\ii \, V(t) \frac{\partial W(t)}{\partial t} \,.
\end{equation}
Multiplication with $V^{-1}(t)$ from the left leads to the equation
\begin{equation}
\ii \, \frac{\partial W(t)}{\partial t} =
h(t) W(t) \,,
\qquad
h(t) = V^{-1}(t) H_0(t) V(t) - \ii V^{-1}(t) \frac{\partial V(t)}{\partial t} \,.
\end{equation}
The matrix representation of this new Hamiltonian~$h(t)$ for general~$F$
as a $(2F+1) \times (2F+1)$
matrix can be written in terms of the following,
somewhat self-explanatory notation,
where the~$2 F+1$~elements on the diagonal are
written in the middle, the~$2F$~entries of the sub-diagonal on the left,
and the~$2F$~entries of the super-diagonal on the right,
\begin{equation}
\label{Hprimedef}
h(t) = \mathrm{diag}
\begin{pmatrix}
\ \ \ \ d_F f(t) \exp \left[ - \ii (2 F -1)\int_{-\tau}^{t} \epsilon_z^2 (t') \,
\dd  t' \right] & 0 &
\ \ \ \ d_F f^*(t) \exp \left[ +\ii (2 F -1) \int_{-\tau}^{t} \epsilon_z^2 (t') \,
\dd  t' \right] \\[1ex]
\noalign{\smallskip}
d_{F-1} f(t) \exp \left[ - \ii (2 F -3) \int_{-\tau}^{t} \epsilon_z^2 (t') \,
\dd  t' \right] & 0 &
d_{F-1} f^*(t) \exp \left[ +\ii (2 F -3) \int_{-\tau}^{t} \epsilon_z^2 (t') \,
\dd  t' \right] \\[1ex]
\vdots & \vdots & \vdots \\ \ \ \ \ d_{F} f(t) \exp \left[ +\ii (2F-1)
\int_{-\tau}^{t} \epsilon_z^2 (t') \, \dd  t' \right] & 0
& \ \ \ \ d_{F} f^*(t) \exp \left[ - \ii (2F-1) \int_{-\tau}^{t} \epsilon_z^2 (t') \,
\dd  t' \right] \\
& 0 & \end{pmatrix} \,.
\end{equation}
\end{widetext}
Here, $f(t)$ is a single complex function that contains all information about the
magnetic fields in the atom's rest frame,
\begin{equation}\label{fdef}
f(t) = \big(\beta_x(t) + \ii \, \beta_y(t) \big) \;
\exp \left( - \ii \int_{-\tau}^{t} \beta_z(t') \, \dd t' \right) \,.
\end{equation}
The coefficients $d_F$ need to be explained.  They are defined as the
corresponding entries (rational numbers and square roots of rational numbers)
of the matrix representation of the~$x$~component of the total angular
momentum~$\vec F$, which is~$F_x = (F_{+} +  F_-)/2$. These entries are
symmetric with respect to the transformation~$M \to -M$, which is why we have
been able to reduce the notation to the quantities~$d_M$ with~$M = 1,\dots, F$
on the sub- and super-diagonals.

A central point of our proposed scheme is to require that the integral over the
square of the cumulative scaled electric field strength seen by the atom along its path
should be equal to an integer multiple $k_{\epsilon}$ of $\pi$,
\begin{equation}\label{muphaselock}
\int_{-\tau}^{\tau} \epsilon_z^2 (t) \, \dd t
\mathop{=}^{!} k_{\epsilon} \, \pi \,.
\end{equation}
The adjustment to an integer value of $k_{\epsilon}$ has been
used in a prototype experiment \cite{AmMuGo2007,Aminithesis2006}.
According to Eq.~\eqref{Vt},
if the evolution of the hyperfine sublevels were exclusively given
by the electric term (no motional or stray magnetic fields), then for
$k_{\epsilon}$ even, the atomic wave function
would return to its initial quantum state if the
quantization condition~\eqref{muphaselock} is fulfilled.

By measuring the temporally constant stray magnetic field
seen by the atom, and by purposely applying a small additional
magnetic field over the electric field plates,
it is possible to also enforce a condition
for the integral of the~$z$~component of the magnetic field seen by the atom
in the experiment,
\begin{equation}\label{bzphaselock}
\int_{-\tau}^{\tau} \beta_z (t) \, \dd t
\mathop{=}^{!} k_{\beta} \, \pi \,.
\end{equation}
Here, it is the assumed that $k_\beta$ is tuned to be integer, and
we note that $k_{\beta} = 0$ is preferred;
this value corresponds to small overall magnetic fields.
Assuming that these adjustments can be achieved perfectly, the diagonal
time-evolution operator~$V(t)$ at the time $t = \tau$
when the atom leaves the fountain takes the simple form
\begin{equation}
\label{ideal}
V (\tau) = \mathrm{diag}
\begin{pmatrix}
 (-1)^{F^2 k_{\epsilon}} \; (-1)^{F k_{\beta}} \\
\noalign{\smallskip}
 (-1)^{(F - 1)^2 k_{\epsilon}} \; (-1)^{(F-1) k_{\beta}} \\ \vdots \\
\noalign{\smallskip}
 \!\!\!\!\!  (-1)^{(-F)^2 k_{\epsilon}} \; (-1)^{-F k_{\beta}} \end{pmatrix} \,,
\end{equation}
which for even $k_{\epsilon}+k_{\beta}$ reduces to the unit matrix.

We express the remainder $W(t)$ of the time evolution operator
as a time-ordered exponential in terms of a Dyson series.
As we are only interested in its effect at the time
$t=\tau$, when the atoms leave the fountain, we write
\begin{equation}
\label{defWseries}
W(\tau) = 1 + \sum_{n=1}^{\infty} w_n \,.
\end{equation}
The terms have the form
\begin{equation}
w_n =
(-\ii)^n \!\!
\int\limits_{-\tau}^{\tau}
\int\limits_{-\tau}^{t_1}
\cdots  \int\limits_{-\tau}^{t_{n-1}} \!\!\!
h(t_1) \, h(t_2) \! \ldots \! h(t_n) \, \dd t_n \ldots \dd t_2\, \dd t_1 \,.
\end{equation}
For the finite time interval $t \in [ -\tau, \tau ]$
considered in this work, the series for $W(t)$ is in fact convergent.

The basic idea of our theoretical analysis is as follows.
The terms $w_n$ describe the deviation of the quantum dynamics
from the idealized result~\eqref{ideal} due to the magnetic field~\eqref{fdef}.
Analytic
estimates of the terms~$w_n$ will allow us to gauge the magnitude and
significance of individual EDM-mimicking
signals with respect to our target sensitivity. An asymptotic expansion of
the terms, based upon a separation of integrands
into fast oscillating electric-field terms and slowly oscillating
magnetic-field terms, leads to those estimates.

The asymptotic expansion is best
illustrated by considering the first order term
\begin{equation}
w_1 =  (- \ii) \int_{-\tau}^{\tau} h(t_1) \, \dd t_1 \,.
\end{equation}
The nonzero entries of the $(2F+1) \times (2F + 1)$
matrix $w_1$ are of the form
\begin{equation}\label{aexbsp1}
J = \int_{-\tau}^{\tau}   f(t) \exp \left( \ii n \!\int_{-\tau}^{t} \!\!\! 
\epsilon_z^2 (t') \, \dd t' \right) \dd t \,,
\end{equation}
where we use $n$ to represent the integers in the exponents in Eq.~\eqref{Hprimedef},
and $f(t)$ is from Eq.~\eqref{fdef}. We envisage
a situation where $|\epsilon_z(t)|^2$ is on the order of a hundred.
The phase in the above integral is therefore highly oscillatory and the integral can be expressed in terms of an
asymptotic expansion, as it is for example explained in Ref.~\cite{BeOr1978}.
This expansion is obtained by repeated integration by parts of Eq.~\eqref{aexbsp1} as
\begin{align}
J =& \biggl[ \frac{\ee^{ \ii n \int_{-\tau}^{t} \!\! \epsilon_z^2 (t') \, \dd t'}}%
{\ii \, n \, \epsilon_z^2 (t)} f(t)  -
 \frac{\ee^{ \ii n \int_{-\tau}^{t} \!\! \epsilon_z^2 (t') \, \dd t'}}%
{[\ii \, n \, \epsilon_z^2 (t)]^2} f'(t) \\[2ex]
&+ \frac{\ee^{ \ii n \int_{-\tau}^{t} \!\! \epsilon_z^2 (t') \, \dd t'}}%
{[\ii \, n \, \epsilon_z^2 (t)]^3}
\left( f''(t) - 2 f'(t) \frac{\epsilon_z'(t)}{\epsilon_z (t)} \right)
\biggr]_{-\tau}^{\tau} 
\!\!\!+ \ldots \nonumber \,.
\end{align}
This equation expresses the time integral over an
arbitrary integrand function $f(t)$
as a function of surface terms which need to be evaluated at the upper and lower limits
of the time interval within the fountain.
In our case, $f(t)$ is given by Eq.~\eqref{fdef} and describes the dependence
on the magnetic field seen by the atom as a function of the scaled time $t$.
The expansion allows us to integrate out the fast oscillations
due to the applied strong electric field and
to concentrate, in the formulation of the quantum dynamics, on the
dependence due to the residual magnetic fields (static and motional).

We can further simplify the expression for the expansion by noting that the exponential
\begin{equation}
\exp \left[ \,\ii\! \int_{-\tau}^{t} \epsilon_z^2 (t') \, \dd  t' \right] \,,
\end{equation}
evaluated at $t=-\tau$, assumes the value of unity.
In all calculations where we employ this expansion,
the integer $n$ in the exponentials in~Eq.~\eqref{Hprimedef}, which has one of
the values $2F-1$, $2F-3, \ldots$, $-(2F-1)$, is odd.
With the adjustment of the integral in
Eq.~\eqref{muphaselock}, the exponential can be evaluated at $t=\tau$ to be
\begin{equation}
\exp \left[ \,\ii\,  n \! \int_{-\tau}^{\tau} \epsilon_z^2 (t') \, \dd t' \right] = \ee^{\ii n k_{\epsilon} \pi} = (-1)^{k_{\epsilon}} \,.
\end{equation}
If in addition we assume the electric field to be constant,
the formula for the asymptotic expansion of
the integral~$J$ then is written as
\begin{equation}\label{asymptoticJ}
\begin{split}
J = & \; \int_{-\tau}^{\tau} f(t) \exp \left[ \,\ii \, n \! \int_{-\tau}^{t} \!\!\! \epsilon_z^2 (t') \, \dd t' \right] \dd t \\
=&- \frac{\ii}{n \mu} \left( (-1)^{k_{\epsilon}}  f(\tau) -f(-\tau) \right) \\
&+ \frac{1}{n^2 \mu^2} \! \left( (-1)^{k_{\epsilon}} f'(\tau) - f'(-\tau) \right) \\
&+ \frac{\ii}{n^3 \mu^3} \left( (-1)^{k_{\epsilon}} f''(\tau) - f''(-\tau)  \right) + {\mathcal O}(\mu^{-4}) \,.
\end{split}
\end{equation}
Such an expansion provides us with a tool to analyze the relative magnitude of
the relevant physical effects.  In our units used in this work the
parameter~$\mu$ for cesium is about $\mu \approx 120$, and for the envisaged
sensitivity of an electron EDM the dimensionless EDM strength in
Eq.~\eqref{Htt} is $\sigma_F = 4 \times 10^{-9}$.  Therefore, all EDM-mimicking
effects of order~$1/\mu^4 \approx 4\cdot 10^{-9}$ have to be considered and
eliminated in order to reach the target accuracy.  In practice the actual
coefficients of terms of order $1/\mu^4$ are small enough that their
contribution is negligible, and for a suitable observable all effects of order
$1/\mu^n$ for $n$ any odd integer can be proved to cancel, leaving only two
terms of order $1/\mu^2$ to compute and to control.  Using asymptotic
expansions like that of Eq.~\eqref{asymptoticJ} for the integrals appearing in
$W(\tau)$, we are able to express the time-evolution operator
\begin{equation}
U_0(\tau) = V(\tau) \, W(\tau) =
{\bm T} \exp \left(-\ii \int_{-\tau}^\tau H_0(t') \dd t' \right)
\end{equation}
when the electric field is constant
in terms of an asymptotic series in inverse powers of $\mu$,
which has the structure
\begin{align}\label{FormalExp}
& U_0(\tau) = 1
\\[2ex]
& + \sum_{i = 1}^\infty \! \frac{1}{ \mu^i } \,
\mathcal{T}_i \left(
\left\{ \int_{-\tau}^\tau \!
\left[ f^{(n)}(t) \right]^m \! \left[ f^{*(k)}(\tau) \right]^\ell \! \dd t
\right\}_{\!n,m,k,\ell} \right) \,,
\nonumber
\end{align}
where the~$\mathcal{T}_i$ are complex $(2F+1) \times (2F+1)$ dimensional coefficient
functions that can be expressed in terms of integrals
of the $n$th derivatives of the stray magnetic field function~$f(t)$
or its complex conjugate $f^*(t)$, and in terms of powers thereof.
This functional dependence is schematically indicated in Eq.~\eqref{FormalExp}
using the curly brackets and the multi-index $\{ n,m,k,\ell \}$.
We have carried out the calculation to fourth order in $1/\mu$,
for $F= 3$, $F=4$, and $F=5$. The individual expressions are too long to include in this
paper, but the calculation is entirely based on the asymptotic expansion
technique described herein.

The expansion of Eq.~\eqref{FormalExp} has applicability beyond the case when the
electric field is constant.  Provided $\epsilon_z(t)$ is approximately constant
on $[-\tau,\tau]$ there exists a transformation of the time variable that
produces an equivalent problem where the electric field is exactly constant and
where the unitary transformation is exactly of the form of Eq.~\eqref{FormalExp},
with a perturbed function $f$.  As far as the cancellation of systematic effects
in an electron EDM experiment, nothing has changed; only the numerical values
of surviving systematics are perturbed.

Before we use this expansion to evaluate the specific terms
of the time-evolution operator, we want to take the time and
discuss the observable to be used in the experiment \cite{AmMuGo2007,Aminithesis2006}.
The choice of the observable
also identifies the EDM-mimicking effects.

%
% Defining the observable
%
\subsection{Defining the observable}

Without magnetic fields in the $x$ and $y$ direction,
the only effect of the movement through the fountain for the atoms would
be a rotation of the
quantum states by a complex phase [see Eqs.~\eqref{Vt} and~\eqref{ideal}].
This follows because
the phase adjustment of the magnetic field integral as well
as the integral of the square of the electric field
relevant to the Stark shift, sets the
complex phase equal to a multiple of $\pi$.
For $k_{\epsilon} + k_{\beta}$ even,
the complex phase is a multiple of $2 \pi$ and
so the atoms are precisely rotated back into the initial state
when they leave the fountain.
The difference between the Hamiltonian
with the EDM term given in Eq.~\eqref{Htt}
and the Hamiltonian without the EDM term given in Eq.~\eqref{H0}
is that the presence of an EDM leads to a small additional
rotation around the $z$ axis, and so the complex phase
would no longer be zero (or equal to an integer multiple of $2\pi$).
Therefore we need an observable
that is sensitive to rotations about the $z$ axis.

In the case of ${}^{133}$Cs, the valence electron is in the $6 S_{1/2}$ state,
so the atom can be in either of the two hyperfine levels~$F=3$ or~$F=4$.
The atoms are prepared in a state~$\left| 4, M \right>$.
After the atoms leave the fountain,
the number of remaining atoms in the~$F=4$ hyperfine level
is measured and compared the total number of atoms.
Some atoms can also transition
into the~$F=3$ hyperfine level of the $6 S_{1/2}$ state.

Let us denote the state in the $F=I+1/2$ hyperfine level in which the atom is
prepared before it enters the electric field as~$\big|\Phi_0\big>$.  After the
atom exits the electric field, it is analyzed by optical pumping with a laser
propagating in the~$z$~direction and tuned to a~$P$~state with spin $F'=F$,
where the laser is linearly polarized. We are interrogating 
quantum transitions with respect to quantum states whose 
quantization $x$ and $y$ axes are tilted by a rotation an angle~$\theta$ 
with respect to the $z$~axis. The states on which we are
projecting thus are the states $R_z(\theta)\big|F, M\big>$, where $R_z(\theta)$ is a
suitable rotation operator.
In this paper, we use the notation $R_u(\alpha)$ to indicate an operator that rotates a
state on which it acts (active representation) about the axis denoted by $u$
and by an angle $\alpha$ that is positive if the rotation is in a positive sense
about the axis $u$ as determined by the usual right-hand rule.
Under these assumptions, we have $R_z(\theta)\big|F, M\big> = 
\exp(-\ii M \, \theta) \, \big|F, M\big>$.
The probability that the atom will be
found in the dark state of the upper hyperfine level is our basic signal and is
given by
\begin{equation}
\label{defRz}
{\mathcal S}(\theta) =
\sum_{M=-F}^F \C_M \left| \left< F M \left| R_z (\theta) U(\tau) \;
\right| \Phi_0 \right> \right|^2.
\end{equation}
where the constants $\C_M$ can be chosen to be any of the sets in
Tables~\ref{table1} or Table~\ref{table2}. Here, $U(\tau)$ is
the time-evolution operator of the full Hamiltonian~\eqref{Htt},
\begin{equation}
U(\tau) = {\bm T} \exp\left( -\ii  \int_{-\tau}^\tau  H(t) \, \dd t \right) \,.
\end{equation}
A small rotation of the atomic states by an EDM can then be detected by taking two
measurements at $\pm \theta$ and
forming the difference. Thus, the observable $P(\theta)$ is given by the equation
\begin{equation}
\label{defP}
\begin{split}
P(\theta) &= \mathcal{S}(\theta) - \mathcal{S}(-\theta) \\
&=
\!\!\! \sum_{M=-F}^F \!\!\! \C_M \!
\left( \left| \left< F M \left| R_z (+\theta) U(\tau)
\right| \Phi_0 \right> \right|^2 \right. \\
& \qquad\quad\ \ \ \, \left.
- \left| \left< F M \left| R_z (-\theta) U(\tau) \right| \Phi_0 \right> \right|^2 \right) \,,
\end{split}
\end{equation}
where because we are using a linearly polarized laser the probabilities~$\C_M$
have the symmetry $\C_M=\C_{-M}$. This observable is
measured again with the direction of the electric field reversed.
The resulting difference is only sensitive to effects that, like that of an EDM, are odd
under the reversal of the electric field direction.

The currently most promising alkali atoms for an actual experiment are cesium
and francium. For cesium, the natural isotope ${}^{133}{\rm Cs}$ is suited
best. It has nuclear spin $I=7/2$, and so the ground state $6 S_{1/2}$ has the
two hyperfine levels~${F=3}$ and~${F=4}$. The proposed detection scheme
requires to use the energetically higher lying state $F=4$.  For francium the
enhancement factor~$R$ is about nine times larger than in cesium, making this
atom particularly attractive.  The francium isotope~${}^{221}$Fr, which
has~$I=5/2$, can be obtained from actinium sources and the dynamics of the
$F=3$ hyperfine level would be investigated there. Due to a higher nuclear
spin~$I=9/2$, leading to a higher total angular momentum~$F=5$ to be used for
the dynamics, the francium isotope~${}^{211}$Fr, is currently the most
promising atom to study.  It can be obtained from accelerator sources, such as
ISOLDE and TRIUMF, and its half-life of $3$~minutes is sufficiently long for a
practical experiment. Yields of francium at existing and planned facilities
(CERN ISOLDE and TRIUMF) and a discussion of the transfer mechanism from the
beam to the magneto-optical trap can be found in
Refs.~\cite{ISOLDE_fr,TRIUMF_fr,TRIUMF_plan,PROJECTX,AuGoOrSp2003,MeEtAl2004}.

In the calculation of~$P(\theta)$,
it is possible to perform a basis transformation which leads to
a number of simplifications.
We define as our new basis states
the following superpositions of the sublevels~$\pm M$, which are obtained from the
original basis states~$|F M \rangle$ by a rotation
about an angle~$\pi/2$ around  the~$y$~axis,
\begin{align}
| s_M \rangle &= \frac{1}{\sqrt{2}}\, R_y (\pi/2) \left( \left| F M \right> +
\left| F \; -M \right> \right) \,, \\
| a_M \rangle &= \frac{1}{\sqrt{2}}\, R_y (\pi/2) \left( \left| F M \right> -
\left| F \; -M \right> \right) \,.
\end{align}
Here $R_y (\pi/2)$ is the corresponding rotation operator,
and we index the new states by their dependence on~$M$ and suppress the
subscript~$F$. These new states may also be used for the projections carried out
in the calculation of~$P(\theta)$.

The new states have useful symmetries under two transformations.
Let us define the matrix~$B$, which has the entries
\begin{equation}\label{Bdef}
B_{i j } = (-1)^{i+1} \, \delta_{i,j}  \,,
\end{equation}
where the indices run from~$1$ to~$2F+1$.
This matrix is trivially generalized to~$F=3$, $4$, and~$5$:
the $B$ matrix is
a diagonal
$(2F+1) \times (2F+1)$ matrix where the first matrix elements
on the diagonal run $+1,-1,+1,\dots$.
The states~$| s_M \rangle$ are eigenstates of~$B$ with eigenvalue~$+1$, while
the states~$| a_M \rangle$ are eigenstates of~$B$ with eigenvalue~$-1$.

The new states also have a
symmetry under exchange of magnetic quantum
numbers from~$M$ to~$-M$. This exchange can be described by transformation by a second
$(2F+1) \times (2F+1)$ matrix $S$ that has entries
\begin{equation}\label{Sdef}
S_{i j} = \delta_{i, (2F+1)-j}  \,.
\end{equation}
Both $| s_M \rangle$ and~$| a_M \rangle$ are eigenstates of~$S$
with eigenvalues~$(-1)^{F+M}$.

In terms of these new states, and in the case where $p_M=p_{-M}$, we can rewrite $P(\theta)$ as
\begin{equation}
\label{Prewrite}
\begin{split}
P(\theta)= &\sum_{M=0}^{F} \C_M \left( \left| \left< s_M \left| 
R_z (\theta) U(\tau) \right| \Phi_0 \right> \right|^2 \right. \\
& \qquad\quad  \left. - \left| \left< s_M \left| 
R_z^\dagger (\theta) U(\tau) \right| \Phi_0 \right> \right|^2 \right) \\
+{}&\sum_{M=1}^{F} \C_M \left( 
\left| \left< a_M \left| R_z (\theta) U(\tau) \right| \Phi_0 \right> \right|^2 \right. \\
& \qquad\quad \left. -\left| 
\left< a_M \left| R_z^\dagger (\theta) U(\tau) \right| \Phi_0 \right> \right|^2 \right) \,,
\end{split}
\end{equation}
where we have used the fact
that $R_z (-\theta) = R_z^\dagger (\theta)$.

The great utility of the inversion symmetry becomes clear in the next step, as
it allows us to express the difference of the measurements at $\pm \theta$
instead in a simpler, product form [see
Eq.~\eqref{prodform} below].  We define the ``even'' part of a matrix~$A$ as
\begin{equation*}
O_{+} (A) = \half ( A + S A S ) \,,
\end{equation*}
which is even under conjugation with $S$,
\begin{equation*}
S O_{+} (A) S = \half S ( A + S A S )S = O_{+} (A) \,,
\end{equation*}
because $SS=1$. The ``odd'' part
\begin{equation*}
O_{-} (A) = \half ( A - S A S ) \,
\end{equation*}
is odd under conjugation with $S$,
\begin{equation*}
S O_{-} (A) S = \half S ( A - S A S )S =- O_{-} (A) \,.
\end{equation*}
Due to the identity [we use $R_z \equiv R_z(\theta)$]
\begin{widetext}
\begin{equation}
\label{productobservable}
\bigl| \langle \psi_2 \left| R_z A \bigr| \psi_1 \rangle \right|^2 - \left| \langle \psi_2
\left| R_z^\dagger A \right| \psi_1 \rangle \right|^2 
= 
-4 \mathrm{Re} \langle \psi_1 \left| O_{-} (A^\dagger) R_z \right| \psi_2 \rangle \langle \psi_2
\left| R_z^\dagger O_{+} (A) \right| \psi_1 \rangle \,,
\end{equation}
we have the following expression for $P(\theta)$,
\begin{equation}
\label{prodform}
\begin{split}
P(\theta) = &-4\, \mathrm{Re} \biggl[ \sum_{M=0}^{F} \C_M
\left< \Phi_0 \left| O_{-} ( U^\dagger(\tau) ) \right| R_z^\dagger (\theta) s_M \right>
\left< R_z (\theta) s_M \left| O_{+} ( U(\tau)) \right| \Phi_0 \right> \biggr] \\
& -4\, \mathrm{Re} \biggl[ \sum_{M=1}^{F}
\C_M\left< \Phi_0 \left| O_{-} ( U^\dagger(\tau) )
\right| R_z^\dagger (\theta) a_M \right>
\left< R_z (\theta) a_M \left|
O_{+} ( U(\tau)) \right| \Phi_0 \right> \biggr] \,.
\end{split}
\end{equation}
This is an improvement on Eq.~\eqref{Prewrite}
in that we have expressed the
difference between the measurements at~$\pm \theta$ as the real part of
a product: of the part of the time-evolution operator evaluated at $t = \tau$ that is even under
conjugation with~$S$; and a part which is odd under conjugation
with~$S$. In addition, we have separated the projection onto intermediate states
that are even and odd under conjugation with~$B$, namely
$|s_M\rangle$ and $|a_M\rangle$, respectively.

For  $k_{\epsilon}+k_{\beta}$ even, we can use $W(\tau)$ instead
of $U(\tau)$ because $V(\tau)$ is the identity matrix. For $k_{\epsilon}+k_{\beta}$
odd, the effect of $V(\tau)$ is a minus sign for each of the matrix elements
containing $a_M$ for $F=4$, which as we will see later do not contribute
to an EDM or EDM-mimicking signal. We are therefore safe to just use
$W(\tau)$ instead $U(\tau)$. Using Eq.~\eqref{defWseries}, $P(\theta)$
can now be rewritten as
\begin{equation}\label{Pstartformula}
\begin{split}
- \frac{P(\theta)}{4} = \mathrm{Re} \biggl[\, &\sum_{M=0}^{F} \!
\,\C_M  \biggl< \! \Phi_0 \biggl|
O_{-} \biggl(\, \sum_{n=1}^\infty w_n \biggr)^{\!\!\dagger} \biggr|  R_z^\dagger (\theta) s_M \!\! \biggr>
\biggl< R_z (\theta) s_M \biggl|
O_{+} \biggl( 1 + \sum_{n=1}^\infty w_n\biggr) \biggr| \Phi_0 \biggr> \\[2ex]
 +{}& \sum_{M=1}^{F} \C_M\biggl<\! \Phi_0 \biggl|
O_{-} \biggl( \sum_n w_n\biggr)^{\!\!\dagger} \biggr| R_z^\dagger (\theta) a_M\!\! \biggr>
\biggl< R_z (\theta) a_M \biggl| O_{+} \biggl( 1 + \sum_n w_n\biggr)
\biggr| \Phi_0 \biggr>\, \biggr] \,.
\end{split}
\end{equation}
Here, use is made of the fact that $O_{-} (1) =0$. The
observable now being known, we can employ the asymptotic expansion to determine the matrices
$O_{\pm} (w_n)$ which describe the time-evolution of the atoms through the
fountain to identify EDM-mimicking effects. We begin by determining the terms
in the first order of the Dyson series.

%
% First term of the expansion of $\bm W$
%
\subsection{First term of the expansion of $\bm W$}
\label{w1terms}

We continue to use the Hamiltonian $h(t)$,
which does not contain a possible EDM, in order
to isolate all effects that could mimic the presence of an EDM.
We obtain the unitary transformation accurate to order~$1/\mu^2$,
for which as we shall see it is necessary to consider some terms at fourth order
in time-ordered perturbation
theory.  Knowing the terms in the unitary transformation to order~$1/\mu^2$
proves to be sufficient to describe all
EDM-mimicking effects in a suitable observable with error of order~$1/\mu^4$.
The calculational scheme for our observable~$P(\theta)$
implies that we have to determine the $S$-even part of
the matrix~$w_1$, denoted as~$O_{+} (w_1)$,
and the $S$-odd part, denoted as~$O_{-} (w_1)$, separately.
We present these characteristic calculations in some detail and start
by giving the matrix for~$O_{+} (w_1)$, which for general~$F$ has the form
\begin{equation}
O_{+}(w_1) \!=\! \mathrm{diag} \!
\begin{pmatrix}
\ \ \ \ \,- \ii d_F \!\! \int \limits_{-\tau}^{\tau} \!\! \mathrm{Re} \! \left[ f(t) \right] \exp
\Bigl[ - \ii (2 F \!-\!1) \!\! \int \limits_{-\tau}^{t} \!\! \epsilon_z^2 (t') \dd t' \Bigr]  \dd t & 0 &
\ \ \ \  - \ii d_F \!\! \int \limits_{-\tau}^{\tau} \!\! \mathrm{Re} \! \left[ f(t) \right] \exp
 \Bigl[ +\ii (2 F\! -\!1) \!\! \int \limits_{-\tau}^{t} \!\! \epsilon_z^2 (t') \dd t' \Bigr]  \dd t \\
\noalign{\smallskip}
  - \ii d_{F-1} \!\! \int \limits_{-\tau}^{\tau} \!\! \mathrm{Re} \! \left[ f(t) \right] \exp
  \Bigl[ - \ii (2 F \!-\!3) \!\! \int \limits_{-\tau}^{t} \!\! \epsilon_z^2 (t') \dd t' \Bigr]  \dd t & 0 &
   - \ii d_{F-1} \!\! \int \limits_{-\tau}^{\tau} \!\! \mathrm{Re} \! \left[ f(t) \right] \exp
   \Bigl[ +\ii (2 F \!-\!3) \!\! \int \limits_{-\tau}^{t} \!\! \epsilon_z^2 (t') \dd t' \Bigr]  \dd t \\
    \vdots & \vdots & \vdots \\
\ \ \ \ \,- \ii d_{F} \!\! \int \limits_{-\tau}^{\tau} \!\! \mathrm{Re} \! \left[ f(t) \right] \exp
\Bigl[ +\ii (2F\!-\!1) \!\! \int \limits_{-\tau}^{t} \!\! \epsilon_z^2 (t') \dd t' \Bigr] \dd t & 0 &
\ \ \ \ \,- \ii d_{F} \!\! \int \limits_{-\tau}^{\tau} \!\! \mathrm{Re} \! \left[ f(t) \right] \exp
\Bigl[ - \ii (2F\!-\!1) \!\! \int \limits_{-\tau}^{t} \!\! \epsilon_z^2 (t') \dd t' \Bigr] \dd t \\ & 0 &
\end{pmatrix} \,,
\end{equation}
where the $d_F$ are the corresponding entries in the matrix~$F_x$.
A simple calculation reveals that~$S w_1 S$ can be obtained out
of~$w_1$ by exchanging~$f(t)$ and~$f^*(t)$.
The entries of~$O_{+}(w_1)$ are thus $f(t) + f^*(t)$,
which is twice the real part of~$f(t)$.

Similarly, in $O_{-}$, the difference of the function~$f(t)$ from Eq.~\eqref{fdef}
and its complex conjugate gives the imaginary part of~$f(t)$.
There is an additional minus sign for the super diagonal because of the
different order of~$f$ and~$f^*$. As a result of the calculation, $O_{-}(w_1)$
is found to be
\begin{equation}
O_{-}(w_1) = \mathrm{diag}
\begin{pmatrix}
\;\;\;d_F \!\! \int \limits_{-\tau}^{\tau} \! \mathrm{Im} \! \left[ f(t) \right] \exp
\Bigl[  - \ii (2F\!-\!1) \!\! \int \limits_{-\tau}^{t} \!\! \epsilon_z^2 (t') \dd t' \Bigr]  \dd t & \;0 &
\;\;\;\,- d_F \!\! \int \limits_{-\tau}^{\tau} \! \mathrm{Im} \! \left[ f(t) \right] \exp
\Bigl[ +\ii (2F\!-\!1) \!\! \int \limits_{-\tau}^{t} \! \epsilon_z^2 (t') \dd t' \Bigr]  \dd t \\
\noalign{\smallskip}
d_{F-1} \!\! \int \limits_{-\tau}^{\tau} \! \mathrm{Im} \! \left[ f(t) \right] \exp
\Bigl[ - \ii (2 F \!-\!3) \!\! \int \limits_{-\tau}^{t} \! \epsilon_z^2 (t') \dd t' \Bigr]  \dd t & \;0 &
- d_{F-1} \!\! \int \limits_{-\tau}^{\tau} \! \mathrm{Im} \! \left[ f(t) \right] \exp
 \Bigl[ +\ii (2 F \!-\!3) \!\! \int \limits_{-\tau}^{t} \! \epsilon_z^2 (t') \dd t' \Bigr]  \dd t \\
 \vdots & \;\vdots & \vdots \\
\;\;\;\;d_{F} \!\! \int \limits_{-\tau}^{\tau} \! \mathrm{Im} \! \left[ f(t) \right] \exp
\Bigl[ +\ii (2F\!-\!1) \!\! \int \limits_{-\tau}^{t} \! \epsilon_z^2 (t') \dd t' \Bigr] \dd t & \;0 &
\;\;\;\,-d_{F} \!\! \int \limits_{-\tau}^{\tau} \! \mathrm{Im} \! \left[ f(t) \right] \exp
\Bigl[ - \ii (2F\!-\!1) \!\! \int \limits_{-\tau}^{t} \! \epsilon_z^2 (t') \dd t' \Bigr]  \dd t \\
 & \;0 & \end{pmatrix} \,.
\end{equation}
\end{widetext}
If $n=-(2F-1)$, $\dots$, $2F-1$ denotes the prefactor in the
exponentials of the matrix elements of~$O_{+} (w_1)$,
then when~$\epsilon_z$ is constant we find by asymptotic
expansion with error $\mathcal{O}(\mu^{-3})$ we have
\begin{equation}
\begin{split}
 \left[ O_{+}(w_1) \right]_n = \frac{d_j}{n} \frac{1}{\mu} &\left[ \mathrm{Re} [f (-\tau)] - \mathrm{Re} [f(\tau)] (-1)^{k_{\epsilon}} \right] \\
+ \ii \frac{d_j}{n^2} \frac{1}{\mu^2} &\left[ \mathrm{Re} [f' (-\tau)] - \mathrm{Re} [f' (\tau)] \,(-1)^{k_{\epsilon}} \right]
\,.
\end{split}
\end{equation}
The function $f$ is given by
\begin{equation}
f(t) = \left[ \beta_x (t) + \ii \, \beta_y (t) \right] e^{- \ii B(t)} \,,
\end{equation}
with
\begin{equation}
B(t) = \int_{-\tau}^{t}  \beta_z (t') \, \dd t' \,.
\end{equation}
We can simplify results by using the time-symmetry of fields in an atomic
fountain.  Because the atom is assumed to rise and fall along the~$y$~axis, it
passes through the same static magnetic fields when both rising and falling.
In the rest frame of the atom the magnetic fields applied this way are
time-even, {\em i.e.}, even around the time $t=0$.  Similarly the electric field
applied in atom's rest frame, due to the atoms motion through a static electric
field that we have assumed is parallel to the $z$~axis, is time-even.  The
motional magnetic field applied in the atom's rest frame, which arises due to
the Lorentz transformation of the electric field and in our geometry is always
parallel to the~$x$~axis, is time-odd around~${t=0}$ because the atomic
velocity changes sign when the atom falls while the direction of the electric
field does not.

We can split the magnetic field along the~$x$~axis, $\beta_x (t)$, into
its static and time-even part $x_e(t)$ and its motional magnetic and time-odd
part~$x_o (t)$. Thus we have the following facts:
\begin{itemize}
 \item $x_o(t) \equiv \beta_x({\mathrm{motion}}; t)$ \;\; is time-odd,
 \item $x_e (t) \equiv \beta_x({\mathrm{static}}; t)$ \;\; is time-even,
 \item $y_e (t) \equiv \beta_y({\mathrm{static}}; t)$ \;\;  is time-even,
 \item $\epsilon_z (t)$ is time-even around $t=0$,
 \item $z_e (t) \equiv \beta_z({\mathrm{static}}; t)$ is time-even, \\
and in addition
 $B(\tau) = \int_{-\tau}^{\tau}  \beta_z (t') \, d t' = k_{\beta} \pi$.
\end{itemize}
Again specializing to the case of a constant electric field~$\epsilon_z$,
to order $1/\mu$ we find
\begin{align}
& \mathrm{Re} [f (-\tau)] - \mathrm{Re} [f (\tau)] (-1)^{k_{\epsilon}}
=  \mathrm{Re} \left[ \beta_x (-\tau) + \ii \beta_y (-\tau) \right] \nonumber \\
& \qquad - (-1)^{{k_{\epsilon}}+k_{\beta}} \;
\mathrm{Re} \left[ \beta_x (\tau) + \ii \beta_y (\tau) \right]
\\
& \qquad = \beta_x (-\tau) -  (-1)^{{k_{\epsilon}}+k_{\beta}} \beta_x (\tau) \,. \nonumber
\end{align}
Separating the magnetic field in the $x$ direction into its static and motional part and employing the time-symmetries in the ideal fountain, we obtain the result
\begin{align}
& \; \beta_x (-\tau) - (-1)^{k_{\epsilon}+k_{\beta}} \beta_x (\tau) \nonumber \\
& \qquad =
x_e (-\tau) + x_o (-\tau) -
(-1)^{k_{\epsilon}+k_{\beta}} \left[ x_e (\tau) + x_o (\tau) \right] \nonumber \\[1ex]
& \qquad = \left\{ \begin{array}{ll}
2 x_o (-\tau) & \textrm{for} \; k_{\epsilon}+k_{\beta} \; \textrm{even} \,, \\[1ex]
2 x_e (-\tau) & \textrm{for} \; k_{\epsilon}+k_{\beta} \; \textrm{odd} \,.
\end{array} \right.
\end{align}
To get the contribution of order~$1/\mu^2$, we need the derivative
\begin{equation}
\begin{split}
f'(t) &= \frac{\dd}{\dd t} \left[ \beta_x (t) + \ii \beta_y (t) \right] e^{- \ii B(t)} \\
&= \bigl[ \beta_x' (t) + \ii \beta_y' (t) \bigr] e^{- \ii B(t)} \\
& \qquad -\ii \beta_z (t) \bigl[ \beta_x (t) + \ii \beta_y (t) \bigr] e^{- \ii B(t)} \,,
\end{split}
\end{equation}
which allows us to obtain
\begin{equation}
\begin{split}
& \; \mathrm{Re} [f' (-\tau)] - \mathrm{Re} [f' (\tau)] (-1)^{k_{\epsilon}}
=y_e (-\tau) z_e (-\tau) \\
& \qquad - (-1)^{k_{\epsilon}+k_{\beta}} y_e (-\tau) z_e (-\tau) \\
& \qquad + x_e' (-\tau) + x_o' (-\tau) -
(-1)^{k_{\epsilon}+k_{\beta}} x_e' (\tau) x_o' (\tau) \,.
\end{split}
\end{equation}
Because the time derivative an odd function of time is even, 
and the derivative of an even function is odd, we get
\begin{align}
& y_e (-\tau) z_e (-\tau) - (-1)^{k_{\epsilon}+k_{\beta}} y_e (-\tau) z_e (-\tau) \nonumber \\
& \quad + x_e' (-\tau) + x_o' (-\tau) -
(-1)^{k_{\epsilon}+k_{\beta}} x_e' (\tau) x_o' (\tau) \\
& \qquad = \left\{ \begin{array}{ll}
2 x_e' (-\tau) & \textrm{for} \; k_{\epsilon}+k_{\beta} \; \textrm{even} \,, \\[1ex]
2 y_e (-\tau) z_e (-\tau) +2 x_o' (-\tau) & \textrm{for} \;
k_{\epsilon}+k_{\beta} \; \textrm{odd} \,.
\end{array} \right. \nonumber
\end{align}
We also define a standard form for the expansion of the matrices $O_{+} (w_n)$,
which takes the form
\begin{equation}\label{aseriesoplus}
O_{+}(w_n) = \sum_{k=k_0}^\infty \frac{\ii^{n+k}}{\mu^k} \sum_{j}
\mathcal{N}_{n,k}^{(j)} (F) \, \mathcal{G}_{n,k}^{(j)} (\vec{\beta}(t)) \,,
\end{equation}
where the $\mathcal{N}^{(j)}_{n,k}$ are matrices of dimension $(2F+1) \times (2F+1)$
whose entries are real, field-independent constants that depend on~$F$.
The expressions $\mathcal{G}^{(j)}_{n,k}$ are also real and can have
a rather complicated dependence on the
magnetic field $\vec{\beta}(t)$, as well as integrals, derivatives and of powers
thereof, but are independent of~$F$.
The sum over~$j$ is introduced to effect a
natural separation of the terms of a given order
in~$n$ and~$k$ according to specific symmetry properties of the
field-dependent functions~$\mathcal{G}^{(j)}_{n,k}$,
as explained below.

In first order in $n$, where the subscript $n$ denotes the
expansion order for the time-ordered perturbation theory,
only a single term with $j=1$ is required,
and we can suppress the superscript $j$ in our notation;
the sum over $j$ becomes necessary only for the
second and higher orders ($n=2,3,\dots$).
Low-order terms in $n$ will nonetheless contribute to high orders in~$1/\mu$.
Corresponding to each~$n$ is however a
lowest-order nonvanishing contribution in~$1/\mu$. While one's initial guess would be that the sum over the inverse powers
of $1/\mu$ starts at $k=n$, in fact this sum starts at the value $k_0$ defined by
\begin{equation}
\label{k0celing}
k_0 \equiv k_0(n) = \left\{ \begin{array}{cl}
{n}/{2} & \textrm{for} \; n \; \textrm{even,} \\[1ex]
({n+1})/{2} & \textrm{for} \; n \; \textrm{odd} \,,
\end{array} \right.
\end{equation}
or in an alternative notation as the ceiling of~$n/2$
(i.e., the smallest integer larger than or equal to~$n/2$).
This more complicated rule arises from
the cancellations among the various exponential factors in various integrands, as
explained in more detail below.

For $O_{-}(w_n)$, we define an analogous series by
\begin{equation}\label{aseriesominus}
O_{-}(w_n) = \sum_{k=k_0}^\infty \frac{\ii^{n+k+1}}{\mu^k}
\sum_{j} \mathcal{M}^{(j)}_{n,k} (F) \,
\mathcal{B}^{(j)}_{n,k} (\vec{\beta}(t)) \,,
\end{equation}
where we now use $\mathcal{M}$ to denote the respective matrices of real field-independent constants;
$\mathcal{B}$ for the respective magnetic-field dependent, real expressions; and where $k_0$
is as defined in Eq.~\eqref{k0celing}.  The matrices~$\mathcal{M}$ and~$\mathcal{N}$ as well as the expressions $\mathcal{G}$
and $\mathcal{B}$ can be shown to be real to all orders.  Because the only imaginary factors in
Eqs.~\eqref{aseriesoplus} and \eqref{aseriesominus} are the explicit powers of $\ii$, it is straightforward later
to take the real part of the observable~$P(\theta)$.

Suppressing for simplicity of notation the
magnetic field dependence
of $\mathcal{G}^{(j)}_{n,k}$, the leading contributions to~$O_{+} (w_1)$ can be written
\begin{equation}
\begin{split}
O_{+}(w_1) &= - \frac{1}{\mu} \mathcal{N}_{1,1} (F) \;
\mathcal{G}_{1,1} \\
& \qquad - \frac{\ii}{\mu^2} \mathcal{N}_{1,2} (F) \; \mathcal{G}_{1,2}
+ \mathcal{O} (\mu^{-3})  \,,
\end{split}
\end{equation}
with the functions
\begin{equation}
\mathcal{G}_{1,1}  = \left\{ \begin{array}{ll}
2 x_o (-\tau) & \textrm{for} \; k_{\epsilon}+k_{\beta} \; \textrm{even} , \\[1ex]
2 x_e (-\tau) & \textrm{for} \; k_{\epsilon}+k_{\beta} \; \textrm{odd} ,
\end{array} \right.
\end{equation}
and
\begin{equation}
\mathcal{G}_{1,2} \! = \! \left\{ \begin{array}{ll}
\!\! 2 x_e' (-\tau) & \textrm{for} \; k_{\epsilon}\!\!+\!k_{\beta} \,
\textrm{even} , \\[1ex]
\!\! 2 y_e (-\tau) z_e (-\tau) \!+\!
2 x_o' (-\tau) & \textrm{for} \; k_{\epsilon}\!\!+\!k_{\beta} \, \textrm{odd} .
\end{array} \right.
\end{equation}
The coefficient matrices $\mathcal{N}$ for cesium with $F=4$ have the form
\begin{equation}
\mathcal{N}_{1,1} (4) = - \mathrm{diag}
\begin{pmatrix}
- {\sqrt{2}}/{7} & 0 & \V{\sqrt{2}}/{7} \\ - {\sqrt{14}}/{10} & 0 & \V{\sqrt{14}}/{10} \\ - {\sqrt{2}}/{2} & 0 & \V{\sqrt{2}}/{2} \\ - \sqrt{5} & 0 & \V\sqrt{5} \\ \V\sqrt{5} & 0 & -\sqrt{5} \\  \V{\sqrt{2}}/{2} & 0 & -{\sqrt{2}}/{2} \\ \V{\sqrt{14}}/{10} & 0 & -{\sqrt{14}}/{10}  \\ \V{\sqrt{2}}/{7} & 0 & -{\sqrt{2}}/{7} \\ & 0 & \end{pmatrix} \,,
\end{equation}
and
\begin{equation}
\mathcal{N}_{1,2} (4) = - \mathrm{diag}
\begin{pmatrix}
{\sqrt{2}}/{49} & 0 & {\sqrt{2}}/{49} \\ {\sqrt{14}}/{50} & 0 & {\sqrt{14}}/{50} \\  {\sqrt{2}}/{6} & 0 & {\sqrt{2}}/{6} \\  \sqrt{5} & 0 & \sqrt{5} \\ \sqrt{5} & 0 & \sqrt{5} \\  {\sqrt{2}}/{6} & 0 & {\sqrt{2}}/{6} \\ {\sqrt{14}}/{50} & 0 & {\sqrt{14}}/{50}  \\ {\sqrt{2}}/{49} & 0 &{\sqrt{2}}/{49} \\ & 0 & \end{pmatrix}  \,.
\end{equation}

In the notation introduced in Eq.~\eqref{aseriesominus},
we find for $O_{-} (w_1)$
\begin{equation}
\begin{split}
O_{-}(w_1) &= - \frac{\ii}{\mu} \mathcal{M}_{1,1} (F) \, \mathcal{B}_{1,1}  \\
& \qquad + \frac{1}{\mu^2} \mathcal{M}_{1,2} (F) \, \mathcal{B}_{1,2}  + \mathcal{O} (\mu^{-3}) \,.
\end{split}
\end{equation}
The magnetic field dependence is given by
\begin{equation}
\mathcal{B}_{1,1}  = \left\{ \begin{array}{ll}
0  & \textrm{for} \; k_{\epsilon}+k_{\beta} \; \textrm{even} , \\[1ex]
2 y_e (-\tau) & \textrm{for} \; k_{\epsilon}+k_{\beta} \; \textrm{odd} .
\end{array} \right.
\end{equation}
and
\begin{equation}
\mathcal{B}_{1,2} \! = \! \left\{ \begin{array}{ll}
\!\! 2 y_e' (-\tau) \!-\!2 x_o (-\tau) z_e (-\tau) & \textrm{for} \;
k_{\epsilon}\!\!+\!k_{\beta} \; \textrm{even} , \\[1ex]
\!\!-2 x_e (-\tau) z_e (-\tau) & \textrm{for} \;
k_{\epsilon}\!\!+\!k_{\beta} \; \textrm{odd} .
\end{array} \right.
\end{equation}
For cesium with $F=4$, the coefficient matrices are given as
\begin{equation}
\mathcal{M}_{1,1} (4) = -\mathrm{diag}
\begin{pmatrix}
- {\sqrt{2}}/{7} & 0 & -{\sqrt{2}}/{7} \\ - {\sqrt{14}}/{10} & 0 & -{\sqrt{14}}/{10} \\ - {\sqrt{2}}/{2} & 0 & -{\sqrt{2}}/{2} \\ - \sqrt{5} & 0 & -\sqrt{5} \\ \V\sqrt{5} & 0 & \V\sqrt{5} \\  \V{\sqrt{2}}/{2} & 0 & \V{\sqrt{2}}/{2} \\ \V{\sqrt{14}}/{10} & 0 & \V{\sqrt{14}}/{10}  \\ \V{\sqrt{2}}/{7} & 0 & \V{\sqrt{2}}/{7} \\ & 0 & \end{pmatrix} \,,
\end{equation}
and
\begin{equation}
\mathcal{M}_{1,2} (4) = \mathrm{diag}
\begin{pmatrix}
-{\sqrt{2}}/{49} & 0 & {\sqrt{2}}/{49} \\ -{\sqrt{14}}/{50} & 0 & {\sqrt{14}}/{50} \\  -{\sqrt{2}}/{6} & 0 & {\sqrt{2}}/{6} \\  -\sqrt{5} & 0 & \sqrt{5} \\ -\sqrt{5} & 0 & \sqrt{5} \\  -{\sqrt{2}}/{6} & 0 & {\sqrt{2}}/{6} \\ -{\sqrt{14}}/{50} & 0 & {\sqrt{14}}/{50}  \\ -{\sqrt{2}}/{49} & 0 & {\sqrt{2}}/{49} \\ & 0 & \end{pmatrix}  \,.
\end{equation}
For atoms with $F=3$ and $F=5$, the coefficient matrices are given in
Appendices~E and~F of Ref.~\cite{WuMuJe2012app}, respectively.

% 
% HIGHER--ORDER CALCULATION
% 
\section{HIGHER--ORDER CALCULATION}

%
% Second term in the expansion of $\bm W$
%
\subsection{Second term in the expansion of $\bm W$}
\label{w2terms}

In the second order of time-dependent perturbation theory,
we need to investigate the matrix $w_2$,
which takes the more complicated form
\begin{equation}
w_2 =  - \int_{-\tau}^{\tau} \int_{-\tau}^{t_1} h(t_1) h(t_2) \, \dd t_2 \, \dd t_1 \,,
\end{equation}
and where there are more nonzero entries in the resulting matrix. Due to the presence of two
factors in the integrand, each with a phase factor, one's
initial guess would be that the resulting matrices are at least
of order~$1/\mu^2$. However,
terms of lower order in~$1/\mu$ appear because of cancellations in the exponentials. For example, consider
the terms on the diagonal of $O_{+} (w_2)$. The structure of these
integrals is
\begin{equation}\label{w2termform}
\begin{split}
\mathcal{J} =& \int_{-\tau}^{\tau} \int_{-\tau}^{t_1} \biggl[ \ee^{\ii n \left[ E(t_1)- E(t_2)\right]} \\
& \quad \times \left[ f(t_2) f^*(t_1)+ f(t_1) f^*(t_2) \right] \biggr] \, \dd t_2 \, \dd t_1 \,,
\end{split}
\end{equation}
where $n=2F-1,2F-3,\ldots,-(2F-1)$ is an odd integer and we have used
\begin{equation}
E( t ) = \int_{-\tau}^{t} \epsilon_z^2 (t') \, \dd t'
\end{equation}
to shorten the notation. When the electric field is constant, we have
$E(t) = \mu^2(t+\tau)$; and carrying out the asymptotic expansion for the first
integral leads to
\begin{equation}
\begin{split}
\mathcal{J} =& \frac{2\ii}{n \mu} \int_{-\tau}^{\tau} \!\!\! |f (t_1)|^2 \, \dd t_1 + \frac{1}{n^2 \mu^2} \int_{-\tau}^{\tau} \! \frac{\dd}{\dd t_1} |f (t_1)|^2 \, \dd t_1 \\
& + \frac{1}{n^2 \mu^2} \Bigl( 2 |f(-\tau)|^2- f(\tau) f^*(-\tau) (-1)^{k_{\epsilon}} \\
& \qquad\qquad\;\;- f(-\tau) f^*(\tau) (-1)^{k_{\epsilon}} \Bigr) + \mathcal{O} \left( \frac{1}{\mu^3} \right) \,.
\end{split}
\end{equation}
whose lowest contribution is seen to be of~$O(1/\mu)$, not~$O(1/\mu^2)$.
In general the contribution of lowest order in~$1/\mu$ in the expansion
of~$w_n$ is not given by the order in perturbation theory but rather only by the
ceiling of~$n/2$; hence the
introduction of~$k_0$ in Eq.~\eqref{k0celing} and
in the asymptotic series expansions of Eqs.~\eqref{aseriesoplus} and~\eqref{aseriesominus}.

Bringing the expressions for $O_{+} (w_2)$ into the standard
form of~Eq.~\eqref{aseriesoplus} yields
\begin{equation}
\begin{split}
O_{+} (w_2) &= - \frac{\ii}{\mu} \sum_{j=1}^2 \mathcal{N}_{2,1}^{(j)} (F) \,
\mathcal{G}_{2,1}^{(j)}  \\
& \qquad + \frac{1}{\mu^2} \sum_{j=1}^3 \mathcal{N}_{2,2}^{(j)}(F) \,
\mathcal{G}_{2,2}^{(j)}  + \mathcal{O} (\mu^{-3})  \,.
\end{split}
\end{equation}
For the diagonal part of order~$1/\mu$, we have
\begin{equation}
 \mathcal{N}_{2,1}^{(1)} (4) = -\mathrm{diag} \begin{pmatrix}
-2/7 \\
-29/70 \\
-4/5 \\
-7/2 \\
10 \\
-7/2  \\
-4/5  \\
-29/70 \\
 -2/7
\end{pmatrix} \,,
\end{equation}
and for the off-diagonal part we have
\begin{equation}\label{genFmatrix}
 \mathcal{N}_{2,1}^{(2)} (F)  = \begin{pmatrix}
\ddots &  & & & \\
 & 0 & 0 &\dfrac{F(F+1)}{8} & \\[2ex]
 & 0 & 0 &0 & \\[2ex]
 & \dfrac{F(F+1)}{8} & 0 &0 & \\
 &  &  & & \ddots
\end{pmatrix} \,,
\end{equation}
where we use the dots to indicate that all other entries of the
$(2F+1) \times (2F+1)$ matrix $ \mathcal{N}_{2,1}^{(2)} (F)$ are zero.
For the diagonal part,
the magnetic field dependent coefficients are
\begin{equation}
\mathcal{G}_{2,1}^{(1)}  = \int_{-\tau}^{\tau}
\left\{ x_e^2 (t) + x_o^2 (t) + y_e^2 (t) \right\} \, \dd t \,,
\end{equation}
and for the off-diagonal part
\begin{align}
\mathcal{G}_{2,1}^{(2)}  &= 2 (-1)^{k_{\beta}} \!\! \int_{-\tau}^{\tau} \! \biggl\{ \! \left[ x_e^2(t) + x_o^2 (t) -y_e^2 (t) \right] \! \cos g(t) \nonumber \\
& \qquad \qquad\qquad\qquad+2 x_o (t) y_e (t) \sin  g(t)  \biggr\} \, \dd t \,,
\end{align}
where we have defined the new function
\begin{equation}\label{defofphaseg}
g(t) = 2 \int_{0}^t \beta_z (t') \, \dd t' \,,
\end{equation}
which is odd in time.
The diagonal term of order $1/\mu^2$ is
\begin{equation}
 \mathcal{N}_{2,2}^{(1)} (4) = \mathrm{diag} \begin{pmatrix}
-2/49 \\
-443/2450 \\
-16/25 \\
-11/2 \\
-10 \\
-11/2 \\
-16/25 \\
-443/2450 \\
-2/49
\end{pmatrix} \,,
\end{equation}
with
\begin{equation}
\mathcal{G}_{2,2}^{(1)}  = \left\{ \begin{array}{ll}
2 x_o^2 (-\tau) & \textrm{for} \; k_{\epsilon}+k_{\beta} \; \textrm{even} , \\[1ex]
2 x_e^2 (-\tau) + 2 y_e^2 (-\tau) & \textrm{for} \; k_{\epsilon}+k_{\beta} \; \textrm{odd} .
\end{array} \right.
\end{equation}
There is a second term, with superscript $j=2$,
with a coefficient matrix
\begin{equation}
 \mathcal{N}_{2,2}^{(2)} (4) =
 {\displaystyle
 \left(
\begin{array}{c@{\hspace{-0.05cm}}c@{\hspace{-0.05cm}}c@{\hspace{-0.05cm}}c@{\hspace{-0.05cm}}%
c@{\hspace{-0.05cm}}c@{\hspace{-0.05cm}}c@{\hspace{-0.05cm}}c@{\hspace{-0.05cm}}%
c@{\hspace{-0.05cm}}c@{\hspace{-0.05cm}}c}
\sssz& \sssz& \sss {2\sqrt{7}}/{35}\!\!\!\!\! & & & & & & \\[0.5ex]
\sssz& \sssz&\sssz & \sss \!{\sqrt{7}}/{5} & & & & & \\[0.5ex]
\!\!\sss {2\sqrt{7}}/{35} &\sssz &\sssz & \sssz& \sss \sqrt{10} & & & & \\[0.5ex]
 & \sss {\sqrt{7}}/{5}\! &  \sssz& \sssz&\sssz & \sss \!-10 & & & \\[0.5ex]
 & &  \sss \sqrt{10} &\sssz & \sssz&\sssz & \sss \sqrt{10} & & \\[0.5ex]
 & & & \sss \!-10 &\sssz &\sssz &\sssz & \sss {\sqrt{7}}/{5} & \\[0.5ex]
 & & & & \sss \sqrt{10} &\sssz &\sssz & \sssz& \sss {2\sqrt{7}}/{35} \\[0.5ex]
 & & & & & \sss {\sqrt{7}}/{5} & \sssz&\sssz &\sssz \\[0.5ex]
& & & & & & \!\!\!\!\!\sss{2\sqrt{7}}/{35} & \sssz&\sssz
\end{array}
\right)
}
\,.
\end{equation}
Here and in the following we refrain from using the
``diag'' notation for matrices whose structure is beyond tridiagonal.
Blank entries are understood to denote zeros.
The field-dependent terms contain the magnetic fields squared
and so are even when the electric field is reversed,
\begin{equation}
\mathcal{G}_{2,2}^{(2)}  = \left\{ \begin{array}{ll}
x_o^2 (-\tau) & \textrm{for} \; k_{\epsilon}+k_{\beta} \; \textrm{even},  \\[1ex]
x_e^2 (-\tau) - y_e^2 (-\tau) & \textrm{for} \; k_{\epsilon}+k_{\beta} \; \textrm{odd} .
\end{array} \right.
\end{equation}
A third term with $j=3$ is odd in the electric field and given by
\begin{equation}
 \mathcal{N}_{2,2}^{(3)} (4) {=}\,
 {\displaystyle
\left(\!\!\p
\begin{array}{c@{\hspace{-0.05cm}}c@{\hspace{-0.05cm}}c@{\hspace{-0.05cm}}c@{\hspace{-0.05cm}}%
c@{\hspace{-0.05cm}}c@{\hspace{-0.05cm}}c@{\hspace{-0.05cm}}c@{\hspace{-0.05cm}}%
c@{\hspace{-0.05cm}}c@{\hspace{-0.05cm}}c}
\sssz& \sssz& \sss \q\p -{\sqrt{7}}/{105}\m\q\!\!\!\! & & & & & & \\[0.5ex]
\sssz& \sssz&\sssz & \sss \q\p -{\sqrt{7}}/{20}\m\q & & & & & \\[0.5ex]
\sss \q\m{\sqrt{7}}/{105}\p\q &\sssz &\sssz & \sssz& \sss \q\p -{\sqrt{10}}/{2}\m\q & & & & \\[0.5ex]
 & \sss \q\m {\sqrt{7}}/{20}\p\q &  \sssz& \sssz&\sssz & \sss 0 & & & \\[0.5ex]
 & &  \sss \q\m{\sqrt{10}}/{2}\p\q &\sssz & \sssz&\sssz & \sss \q\p\!\! {\sqrt{10}}/{2}\m\q & & \\[0.5ex]
 & & & \sss 0 &\sssz &\sssz &\sssz & \sss \q\p {\sqrt{7}}/{20}\q\m & \\[0.5ex]
 & & & & \sss \q\m -{\sqrt{10}}/{2}\p\q &\sssz &\sssz & \sssz& \sss \q\p{\sqrt{7}}/{105}\q\m \\[0.5ex]
 & & & & & \sss \q\m\!\! -{\sqrt{7}}/{20}\p\q & \sssz&\sssz &\sssz \\[0.5ex]
& & & & & & \sss \q\m\!\!-{\sqrt{7}}/{105}\p\q & \sssz&\sssz
\end{array}
\p\,\right)
}
\,,
\end{equation}
with
\begin{equation}
\mathcal{G}_{2,2}^{(3)}  =  x_e (-\tau) x_o (-\tau) \,.
\end{equation}
For $O_{-}(w_2)$ we find, written in the standard form~\eqref{aseriesominus},
\begin{equation}
\begin{split}
O_{-} (w_2) &= \frac{1}{\mu} \sum_{j=1}^2 \mathcal{M}_{2,1}^{(j)} (F) \,
\mathcal{B}_{2,1}^{(j)}  \\
& \qquad + \frac{\ii}{\mu^2} \sum_{j=1}^4 \mathcal{M}_{2,2}^{(j)} (F) \,
\mathcal{B}_{2,2}^{(j)}  + \mathcal{O} (\mu^{-3})  \,.
\end{split}
\end{equation}
Interestingly, $O_{-} (w_2)$ has no term of order $1/\mu$ on the diagonal.
In the terms on the diagonal, the exponential
factor is the same as the exponential factor of 
Eq.~\eqref{w2termform}, but with the plus sign changed into a minus sign; when the
asymptotic expansion is carried out, the contributions of order~$1/\mu$ now cancel instead of add. Off the diagonal there are still terms
of order~$1/\mu$. These connect only states with $|M|=1$ and the coefficient matrix
has the form
\begin{equation}\label{M211matrix}
\mathcal{M}_{2,1}^{(1)} (F)  = \begin{pmatrix}
\ddots &  & & & \\
 & 0 & 0 & -\dfrac{F(F+1)}{4} & \\[2ex]
 & 0 & 0 &0 & \\[2ex]
 & \dfrac{F(F+1)}{4} & 0 &0 & \\
 &  &  & & \ddots
\end{pmatrix} \,,
\end{equation}
where we use the notation established in Eq.~\eqref{genFmatrix}.
The associated magnetic field dependence is electric-field even
and given by
\begin{equation}
\mathcal{B}_{2,1}^{(1)}  =
2 (-1)^{k_{\beta}} \!\! \int_{-\tau}^{\tau} x_e (t) y_e (t)
\cos  g (t)  \, \dd t \,.
\end{equation}
There also is an electric-field odd contribution with the same coefficient matrix, {\em i.e.}
\begin{equation}
\mathcal{M}_{2,1}^{(2)} (F) = \mathcal{M}_{2,1}^{(1)} (F) \,,
\end{equation}
and with $\mathcal{B}_{2,1}^{(2)}$ given by
\begin{equation}
\mathcal{B}_{2,1}^{(2)} =
- 2 (-1)^{k_{\beta}} \!\! \int_{-\tau}^{\tau} \!\! x_o (t) x_e (t) \sin  g (t)  \, \dd t \,.
\end{equation}
There are four matrices that occur at the order $1/\mu^2$.  We have
\begin{equation}
\mathcal{M}_{2,2}^{(1)} (4) = \mathrm{diag} \begin{pmatrix}
-2/49  \\
-243/2450  \\
-9/25 \\
-9/2 \\
\V 0  \\
\V 9/2 \\
\V 9/25 \\
\V 243/2450 \\
\V 2/49
\end{pmatrix} \,,
\end{equation}
where the magnetic field dependence is a mix of terms that are electric-field even and electric-field odd,
\begin{equation}
\begin{split}
\mathcal{B}_{2,2}^{(1)} &=
- \int_{-\tau}^{\tau} z_e (t) \left[ x_e^2 (t) + x_o^2 (t) + y_e^2 (t) \right] \, \dd t \\
& \quad +\left\{ \begin{array}{ll}
- 2 \int_{-\tau}^{\tau} x_o' (t) y_e(t) \, \dd t & \textrm{for} \; k_{\epsilon}+k_{\beta} \; \textrm{even}, \\[1ex]
+ 2 \int_{-\tau}^{\tau}  x_o (t) y_e'(t) \,\dd t & \textrm{for} \; k_{\epsilon}+k_{\beta} \; \textrm{odd} ,
\end{array} \right.
\end{split}
\end{equation}
and
\begin{equation}
\mathcal{M}_{2,2}^{(2)} (4) \,{=}
 {\displaystyle
 \left(\,\,\,
\begin{array}{c@{\hspace{-0.05cm}}c@{\hspace{-0.05cm}}c@{\hspace{-0.05cm}}c@{\hspace{-0.05cm}}%
c@{\hspace{-0.05cm}}c@{\hspace{-0.05cm}}c@{\hspace{-0.05cm}}c@{\hspace{-0.05cm}}%
c@{\hspace{-0.05cm}}c@{\hspace{-0.05cm}}c}
\sssz& \sssz& \sss \q\p{\sqrt{7}}/{105}\m\q & & & & & & \\[0.5ex]
\sssz& \sssz&\sssz & \sss \q\p{\sqrt{7}}/{20}\m\q & & & & & \\[0.5ex]
\sss \q\m{\sqrt{7}}/{105}\p\q &\sssz &\sssz & \sssz& \sss \q\p{\sqrt{10}}/{2}\m\q & & & & \\[0.5ex]
 & \sss \q\m{\sqrt{7}}/{20}\p\q &  \sssz& \sssz&\sssz & \sss 0 & & & \\[0.5ex]
 & &  \sss \q\m{\sqrt{10}}/{2}\p\q &\sssz & \sssz&\sssz & \sss \q\p-{\sqrt{10}}/{2}\m\q\!\! & & \\[0.5ex]
 & & & \sss 0 &\sssz &\sssz &\sssz & \sss \q\p-{\sqrt{7}}/{20}\q\m\!\! & \\[0.5ex]
 & & & & \sss \q\m -{\sqrt{10}}/{2}\p\q &\sssz &\sssz & \sssz& \sss \q\p \!\!-{\sqrt{7}}/{105}\m\q \\[0.5ex]
 & & & & & \sss \q\m-{\sqrt{7}}/{20}\p\q & \sssz&\sssz &\sssz \\[0.5ex]
& & & & & & \sss \q\m \!\!\!\!-{\sqrt{7}}/{105}\p\q & \sssz&\sssz
\end{array}
\p\,\right)
} \,.
\end{equation}
where the magnetic-field dependence is electric-field odd, only:
\begin{equation}
\mathcal{B}_{2,2}^{(2)}   = x_o (-\tau) y_e (-\tau) \,.
\end{equation}
The next matrix is
\begin{equation}
\mathcal{M}_{2,2}^{(3)} (4) {=}
 {\displaystyle
 \left(\p\!\!
\begin{array}{c@{\hspace{-0.05cm}}c@{\hspace{-0.05cm}}c@{\hspace{-0.05cm}}c@{\hspace{-0.05cm}}%
c@{\hspace{-0.05cm}}c@{\hspace{-0.05cm}}c@{\hspace{-0.05cm}}c@{\hspace{-0.05cm}}%
c@{\hspace{-0.05cm}}c@{\hspace{-0.05cm}}c}
\sssz& \sssz& \sss \q\p-{4 \sqrt{7}}/{35}\m\q\!\!\!\! & & & & & & \\[0.5ex]
\sssz& \sssz&\sssz & \sss \q\p-{2\sqrt{7}}/{5}\m\q & & & & & \\[0.5ex]
\sss \q\m{4\sqrt{7}}/{35}\p\q &\sssz &\sssz & \sssz& \sss \q\p-2 \sqrt{10}\m\q & & & & \\[0.5ex]
 & \sss \q\m{2\sqrt{7}}/{5}\p\q &  \sssz& \sssz&\sssz & \sss 0 & & & \\[0.5ex]
 & &  \q\m\sss 2 \sqrt{10} \p\q&\sssz & \sssz&\sssz & \sss \q\p-2 \sqrt{10}\m\p\!\!\!\! & & \\[0.5ex]
 & & & \sss 0 &\sssz &\sssz &\sssz & \sss \q\p\!\!-{2 \sqrt{7}}/{5}\m\q & \\[0.5ex]
 & & & & \sss \q\m 2 \sqrt{10}\p\q &\sssz &\sssz & \sssz& \sss \q\p\!\!-{4 \sqrt{7}}/{35}\m\q \\[0.5ex]
 & & & & & \sss \q\m{2 \sqrt{7}}/{5}\p\q & \sssz&\sssz &\sssz \\[0.5ex]
& & & & & & \sss  \q\m{4 \sqrt{7}}/{35} \p\q& \sssz&\sssz
\end{array}
\p\right)
} ,
\end{equation}
with the electric-field even dependence
\begin{equation}
\mathcal{B}_{2,2}^{(3)}  = \left\{ \begin{array}{ll}
0 & \textrm{for} \; k_{\epsilon}+k_{\beta} \; \textrm{even}, \\[1ex]
x_e (-\tau) y_e(-\tau) & \textrm{for} \; k_{\epsilon}+k_{\beta} \; \textrm{odd} ;
\end{array} \right.
\end{equation}
and the last matrix, which only acts on states with
$|M|=1$, is given by
\begin{equation}\label{M224matrix}
\mathcal{M}_{2,2}^{(4)} (F) = \begin{pmatrix}
\ddots &  & & & \\
 & 0 & 0 & F(F+1) & \\[2ex]
 & 0 & 0 &0 & \\[2ex]
 & -F(F+1) & 0 &0 & \\
 &  &  & & \ddots
\end{pmatrix} \,,
\end{equation}
where the dependence is electric-field even,
\begin{equation}
\mathcal{B}_{2,2}^{(4)}   = \left\{ \begin{array}{ll}
0 & \textrm{for} \; k_{\epsilon}+k_{\beta} \; \textrm{even} \,, \\[1ex]
x_e (-\tau) y_e(-\tau) & \textrm{for} \; k_{\varepsilon}+k_{\beta} \; \textrm{odd} \,.
\end{array} \right.
\end{equation}
This concludes the contribution from $w_2$.
The terms of order~$1/\mu^2$ that derive from third and fourth order in
perturbation theory are given in Appendices~C, D, G and H
of Ref.~\cite{WuMuJe2012app}, respectively.

%
% Calculation of the observable
%
\subsection{Calculation of the observable}

At this stage, we can keep all terms of order $1/\mu^2$ in
our observable $P(\theta)$ by truncating Eq.~\eqref{Pstartformula} to
\begin{equation}\label{Pgenformula}
\begin{split}
\!\!\!\!\!\!- P(\theta&)/4 = \\
\mathrm{Re} \biggl[ &  \sum_{M=0}^F \C_M \langle \Phi_0  \left| O_{-} (w_1 \!+\! w_2 \!+\! w_3 \!+\! w_4)^\dagger \right| R_z s_M \rangle \\
& \;\;\; \times \langle R_z s_M \left| O_{+} (1\!+\!w_1 \!+\! w_2 \!+\! w_3 \!+\! w_4) \right| \Phi_0 \rangle \\
{}+{}&   \sum_{M=1}^F \C_M \langle \Phi_0 \left| O_{-} (w_1 \!+ w_2 \!+\! w_3 \!+\! w_4)^\dagger \right| R_z a_M \rangle \\
& \;\;\; \times \langle R_z a_M \left| O_{+} (1\!+\!w_1 \!+\! w_2 \!+\! w_3 \!+\! w_4) \right| \Phi_0 \rangle \biggr] \,.
\end{split}
\end{equation}
Simplifications of this expression are possible using the two symmetry
operators $S$ and $B$.
Through these symmetries, the states can be classified into orthogonal subspaces.
Matrix elements between states from different subspaces then vanish due to
the orthogonality. This, however, requires us to specify the initial state and
the total angular momentum. Otherwise, the behavior with respect to the
$S$ and $B$ symmetries cannot be determined.

Consequently, we now specialize to $F=4$ and to one of the initial states
\begin{align}
\left| \Phi_0 \right> &= \frac{1}{\sqrt{2}} \left( \left| 4, 4 \right> + \left| 4, -4 \right> \right) \,, \\
\left| \Phi_0 \right> &= \frac{1}{\sqrt{2}} \left( \left| 4, 2 \right> + \left| 4, -2 \right> \right) \,.
\end{align}
Both of these have the same behavior with respect to the two symmetries $B$ and $S$, {\em i.e.}
\begin{align}
B \left| \Phi_0 \right> &= (+1) \left| \Phi_0 \right> \,, \\
S \left| \Phi_0 \right> &= (+1) \left| \Phi_0 \right>\,.
\end{align}
Let us examine the matrix element in the first line of Eq.~\eqref{Pgenformula}
\begin{equation*}
\langle \underbrace{\Phi_0}_{b=+1} \big|
O_{-} (\underbrace{w_1}_{b=-1} +
\underbrace{w_2}_{b=+1} +
\underbrace{w_3}_{b=-1} +
\underbrace{w_4}_{b=+1})^\dagger \big| \underbrace{R_z s_M}_{b=+1}\rangle \,.
\end{equation*}
and examine how the pieces transform under~$B$.  We see that the state
$O_{-}(w_1)^\dagger \big| R_z s_M \rangle$ belongs to the eigenspace of~$B$
with eigenvalue~$b=-1$, while $\big|\Phi_0 \big>$ has eigenvalue~$+1$, and so the
overlap of these states vanishes. Similarly the overlap involving
$O_{-}(w_3)^\dagger$ vanishes, and so the matrix element as a whole reduces to
\begin{equation}
\to \langle \Phi_0  \left| O_{-} (w_2 + w_4)^\dagger \right| R_z s_M \rangle  \,.
\end{equation}
An analogous analysis for the other matrix elements allows to reduce~$P(\theta)$ to
\begin{equation}
\begin{split}
\!\!- {P(\theta)}/{4} = \mathrm{Re} \biggl[  &\sum_{M=0}^F
\! \C_M\, \langle \Phi_0  \left| O_{-} (w_2 \!+\! w_4)^\dagger \right| R_z s_M \rangle \\
& \;\;\,\, \times \langle R_z s_M \bigl| O_{+} (1 \!+\! w_2 \!+\! w_4)
\bigr| \Phi_0 \rangle \\
{}+{}&  \sum_{M=1}^F \!\,\C_M \langle \Phi_0 \left| O_{-} (w_1 \!+\! w_3)^\dagger
\right| R_z a_M \rangle \\
& \;\;\,\, \times \langle R_z a_M
\bigl| O_{+} ( w_1 \!+\! w_3) \bigr| \Phi_0 \rangle \biggr] \,,
\end{split}
\end{equation}
We now consider the symmetry of the various contributions under~$S$.
The real and the imaginary part of the rotation matrix $R_z$ transform
differently, {\em i.e.} $S\, \mathrm{Re} (R_z) S = + \mathrm{Re} (R_z) $ and
$S\, \mathrm{Im} (R_z) S = - \mathrm{Im} (R_z) $. Recalling that each
of the initial states $\big|\Phi_0\big>$ under consideration have eigenvalue~$+1$ under~$S$,
we can eliminate further terms from $P(\theta)$ and find
\begin{align}
{P(\theta)}/{4} &= \nonumber \\
\mathrm{Re}  \biggl[
- \ii \!\! &\!\!\!\sum_{M =0,2,4} \!\!\! \C_M \langle \Phi_0 \! \left| O_{-} (w_2 \!+\! w_4)^\dagger \right| \! \mathrm{Im} (R_z) s_M \rangle \nonumber \\
& \qquad \times \langle \mathrm{Re} (R_z) s_M \left| O_{+} (1+ w_2 + w_4) \right| \Phi_0 \rangle \nonumber \\
+ \ii \!\! &\sum_{M =1,3} \!\! \C_M \langle \Phi_0  \left| O_{-} (w_2 + w_4)^\dagger \right| \mathrm{Re} (R_z) s_M \rangle \nonumber \\
& \qquad \times  \langle \mathrm{Im} (R_z) s_M \left| O_{+} (1+w_2 + w_4) \right| \Phi_0 \rangle \nonumber \\
- \ii \!\! &\sum_{M =2,4} \!\! \C_M \langle \Phi_0 \left| O_{-} (w_1 + w_3)^\dagger \right| \mathrm{Im} (R_z) a_M \rangle \nonumber \\
& \qquad \times \langle \mathrm{Re} (R_z) a_M \left| O_{+} (w_1 + w_3) \right| \Phi_0 \rangle \label{phaseexplanation} \\
+ \ii \!\! &\sum_{M =1,3} \!\! \C_M \langle \Phi_0 \left| O_{-} (w_1 + w_3)^\dagger \right| \mathrm{Re} (R_z) a_M \rangle \nonumber \\
& \qquad \times \langle \mathrm{Im} (R_z) a_M \left| O_{+} (w_1 + w_3) \right| \Phi_0 \rangle \biggr] \nonumber \,.
\end{align}
The last step in the simplification is to take the real part. Because
the total expression carries a pre-factor of~$\ii$, only those
products of matrix
elements contribute that contain an additional imaginary unit.
From Eq.~\eqref{aseriesoplus},
we recall that the term in order~$1/\mu^m$ of $O_{+} (w_n)$ is imaginary
as~$\ii^{n+m}$ is imaginary, and from Eq.~\eqref{aseriesominus} that the term
in order~$1/\mu^m$ of~$O_{-} (w_n)$ is imaginary as~$\ii^{n+m+1}$ is imaginary.
So, in a product
of these two matrices the total phase is~$\ii^{n_++n_-+m_++m_-+1}$.
Here
the subscripts $+$ and $-$ refer to a contribution from $O_+$ and $O_-$,
respectively, and $n$ refers to the order of a term in perturbation theory
and $m$ refers to the power in $1/\mu$ at which the contribution contributes.
In each of the products of matrix elements in
Eq.~\eqref{phaseexplanation}, the phase $\ii^{n_++n_-}$ is always real
because~$n_+$ and~$n_-$ are either both
odd or both even. The remaining phase~$\ii^{m_++m_-+1}$ is imaginary
only when the order in~$1/\mu$, equal to
$m_++m_-$, is even. Thus, in the final result for~$P(\theta)$ only
even powers of~$1/\mu$ survive.  In fact this result is perfectly general:
the same arguments applied to the full expression of
Eq.~\eqref{Pstartformula} show that the asymptotic expansion for $P(\theta)$ contains only {\sl even\/}
powers of $1/\mu$, for all powers however high.

\begin{figure}[htb]
\begin{center}
\includegraphics[width=0.93\linewidth]{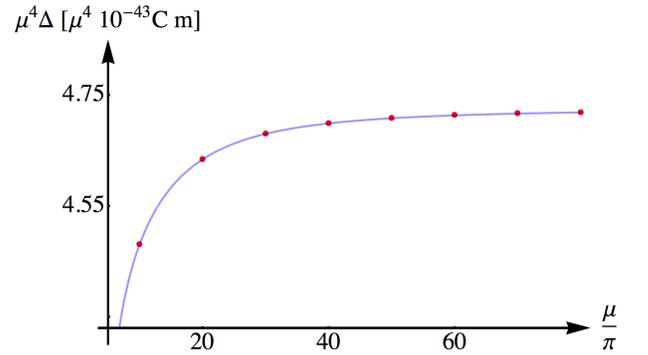}
\end{center}
\caption{\label{figg3} Plot as a function of $\mu$ of the scaled difference
$\Delta= \mu^4(\mathcal{P}^o_{\Delta} - \mathcal{P}^o_{\Delta\, \mathrm{num}})$;
here $\mathcal{P}^o_{\Delta}$ is defined by Eq.~\eqref{plotsignal}, and the
difference is taken between the result
from the asymptotic expansion based on Eq.~\eqref{Po44} and the
result from a numerical calculation $\mathcal{P}^o_{\Delta \, \mathrm{num}}$.
The scaled difference approaches a constant as $\mu \to \infty$ and
thus confirms the error estimate.}
\end{figure}

Therefore though we have kept only terms of order $1/\mu$ and $1/\mu^2$ in the unitary transformation
in writing Eq.~\eqref{Pgenformula}, and therefore while one might expect the error in the resulting
expression for~$P(\theta)$ to be of order $1/\mu^3$, the error is in fact of order $1/\mu^4$.
We confirm this result by comparing the result for the observable in Eq.~\eqref{Po44}
below with a numerical solution for the problem,
as shown in Fig.~\ref{figg3}, where the
difference between the calculation here and a numerical
solution multiplied by~$\mu^4$ is plotted.
The difference converges to a constant
for~$\mu \gg 1$.
Similar tests have been performed to ensure that the expansion of
the time-evolution operator~$U(\tau)$ includes all effects up
to including order~$1/\mu^2$.

At this point we can go back to our full Hamiltonian including an EDM,
which as we recall from Eq.~\eqref{Htt} has the form
\begin{equation}
H(t) = \epsilon_z^2 (t) \, F_z^2 + \vec{\beta}(t) \cdot \vec {F} +
\sigma_F \, \epsilon_z(t) \, F_z \,.
\end{equation}
The electron EDM $d_{\textrm e}$ enters the dimensionless EDM coupling
$\sigma_F = - d_{\textrm e} R/(F \, A_{\rm S} \, E_S)$;
it is assumed to be very small and
we only need to include this effect in first order.
We use a time-evolution operator
for the EDM part of the Hamiltonian in Eq.~\eqref{Htt},
\begin{equation}\label{EDMWt}
\mathcal{W} = 1 -\ii \int_{-\tau}^{\tau} \sigma_F \, \epsilon_z (t) \, F_z \, \dd t \equiv 1
+ \ii \, D \, F_z \,,
\end{equation}
where we have defined
\begin{equation}
\label{defD}
D = -\sigma_F \int_{-\tau}^{\tau} \epsilon_z (t) \, \dd t \,,
\end{equation}
where $\sigma_F$ has been defined in Eq.~\eqref{sigmaF}.
The EDM term adds the $D$ term to  our observable,
\begin{align}
{P(\theta)}&/{4} = \nonumber\\
D \! \biggl[\!\!  &\;\sum_{M =1,3} \!\!\! \C_M \! \left< \Phi_0 \right|
\! F_z \! \left| \mathrm{Re} (R_z) a_M \right> \!\!  \langle \mathrm{Im} (R_z) a_M
| \Phi_0 \rangle \nonumber \\
 - \!\!\!\!\! &\sum_{M =0,2,4} \!\!\!\!\! \C_M \! \left< \Phi_0 \right| \! F_z \!
\left| \mathrm{Im} (R_z) a_M \right>  \! \langle \mathrm{Re} (R_z) a_M
| \Phi_0 \rangle \biggr] \nonumber \\
 - \frac{1}{\mu^2} \! \biggl[ \! &\,\,\sum_{M =1,3} \!\!\! \C_M \! \left< \Phi_0 \right|
\! \sum_{j=1}^3 \!\mathcal{M}_{2,2}^{(j)}(4)\, \mathcal{B}^{(j)}_{2,2} \!
\left| \mathrm{Re} (R_z) a_M \right> \nonumber \\
& \qquad  \times  \langle \mathrm{Im} (R_z) a_M \left| 1 \right| \Phi_0 \rangle \\
  - \!\!\!\! &\sum_{M =0,2,4} \!\!\!\!\! \C_M \!\left< \Phi_0 \right|
\sum_{j=1}^3 \mathcal{M}_{2,2}^{(j)}(4) \mathcal{B}^{(j)}_{2,2}
 \left| \mathrm{Im} (R_z) a_M \right> \nonumber \\
& \qquad \times \langle \mathrm{Re} (R_z) a_M \left| 1 \right| \Phi_0 \rangle \nonumber \\
 + \!\!\!\! &\;\sum_{M =1,3} \!\!\! \C_M \!\left< \Phi_0  \left|  \mathcal{M}_{1,1}(4)
\mathcal{B}_{1,1} \right| \mathrm{Re} (R_z) s_M \right> \nonumber \\
& \qquad \times  \left< \mathrm{Im} (R_z) s_M \left|  \mathcal{N}_{1,1}(4)
\mathcal{G}_{1,1} \right| \Phi_0 \right> \nonumber \\
 - \!\!\!\! &\;\sum_{M =2,4} \!\!\! \C_M \left< \Phi_0  \left| \mathcal{M}_{1,1}(4)
\mathcal{B}_{1,1} \right| \mathrm{Im} (R_z) s_M \right> \nonumber \\
& \qquad \times  \left< \mathrm{Re} (R_z) s_M \left| \mathcal{N}_{1,1}(4)
\mathcal{G}_{1,1} \right| \Phi_0 \right>\biggr] \!+\! \mathcal{O} (\mu^{-4})\nonumber \,.
\end{align}
In this expression, we now take a closer look at the terms and realize that the
important difference between~$\mathcal{B}_{2,2}^{(2)}$
and~$\mathcal{B}_{2,2}^{(3)}$ is that $\mathcal{B}_{2,2}^{(2)}$ reverses sign
when the direction of the externally applied electric field is reversed while
$\mathcal{B}_{2,2}^{(3)}$ stays the same. The signal in which we are interested
is $D$, which also changes sign when the electric field is reversed. So, if we
measure~$P(\theta)$ twice, once~with the electric field along the $z$
direction, and again~with the electric field in the $-z$ direction, and
subtract the results, then all electric field even terms are canceled in
$P(\theta)$. For the electric field odd part of~$P(\theta)$, we~find
\begin{align}
\label{defPo}
\mathcal{P}^o(\theta) ={}& \frac12 \bigl( P(\theta)_{+E} - P(\theta)_{-E} \bigr)  \\
{}={}& 4 \!\!\! \sum_{M =1,3} \!\!\! \C_M \left< \mathrm{Im} (R_z) a_M | \Phi_0 \right>
\left< \Phi_0 \right| D F_z  \nonumber \\
& \!\!\!-\! {\mu^{-2}}  \bigl[ \mathcal{M}_{2,2}^{(1)}(4) \mathcal{B}^{(1,o)}_{2,2} \!+\!
\mathcal{M}_{2,2}^{(2)}(4) \mathcal{B}^{(2)}_{2,2} \bigr]
 \left| \mathrm{Re} (R_z) a_M \right> \nonumber \\
{}-{}& 4 \!\!\!\!\! \sum_{M =0,2,4} \!\!\!\!\! \C_M  \left< \mathrm{Re} (R_z) a_M
| \Phi_0 \right>  \left< \Phi_0 \right| D F_z  \nonumber \\
& \!\!\!-\! {\mu^{-2}} \bigl[ \mathcal{M}_{2,2}^{(1)}(4) \mathcal{B}^{(1,o)}_{2,2} \!+\!
\mathcal{M}_{2,2}^{(2)}(4) \mathcal{B}^{(2)}_{2,2} \bigr]
\left| \mathrm{Im} (R_z) a_M \right> \nonumber  \\
& \quad + \mathcal{O} (\mu^{-4}) \nonumber \,.
\end{align}
For $F=4$, the observable $P(\theta)[\Phi_0]$ corresponding to two different
initial states $\big|\Phi_0\big>$ evaluates to
\begin{align}
&\mathcal{P}^o (\theta)
\left[ \frac{1}{\sqrt{2}} (\left| 4,4 \right> + \left| 4,-4 \right> ) \right]
=\label{Po44} \\
& \sin (8 \theta)  \biggl( \! 4 D \!- \frac{2 \mathcal{B}^{(1,o)}_{2,2}}{49 \mu^2} \! \biggr) \!\frac{1}{32} \bigl( 35 \C_0\!  + 28 \C_2\! + \C_4\! - 56 \C_1\!  - 8 \C_3  \bigr)\nonumber \\
& \; + \frac{1}{15} \frac{\mathcal{B}^{(2)}_{2,2}}{\mu^2} \biggl( \sin (4 \theta) \cos (2 \theta) \left[ \C_1 -\C_3 \right] \nonumber \\
& \qquad\quad\;\, - \frac{1}{4} \sin (2 \theta) \cos (4 \theta) \left[ 5 \C_0 - 4 \C_2 - \C_4 \right] \biggr) + \mathcal{O} (\mu^{-4}) \,, \nonumber \\
&\mathcal{P}^o (\theta)
\left[ \frac{1}{\sqrt{2}} (\left| 4,2 \right> + \left| 4,-2 \right> ) \right]
=  \label{Po42} \\
& \sin (4 \theta) \biggl(\! 2 D - \frac{9 \mathcal{B}^{(1,o)}_{2,2}}{25 \mu^2} \!\biggr)\frac{1}{8} \bigl( 5 \C_0 + 4 \C_2 + 7 \C_4 -2 \C_1  -14 \C_3 \bigr)  \nonumber \\
& \;+ \frac{1}{15} \frac{\mathcal{B}^{(2)}_{2,2}}{\mu^2} \biggl( \sin (2 \theta) \cos (4 \theta) \left[ \C_1 -\C_3 \right] \nonumber \\
& \qquad\quad\;\, - \frac{1}{4} \sin (4 \theta) \cos (2 \theta) \left[ 5 \C_0 - 4 \C_2 - \C_4 \right] \biggr) + \mathcal{O} (\mu^{-4}) \,. \nonumber
\end{align}
Here, the electric-field odd part  $\mathcal{B}^{(1,o)}_{2,2}$ of $\mathcal{B}^{(1)}_{2,2}$ is
\begin{equation}
\label{B1o}
\mathcal{B}^{(1,o)}_{2,2}  = \left\{ \begin{array}{ll}
- 2 \int_{-\tau}^{\tau} x_o' (t) y_e(t) \, \dd t & \textrm{for} \; k_{\epsilon}+k_{\beta} \; \textrm{even} \\[1ex]
+ 2 \int_{-\tau}^{\tau}  x_o (t) y_e'(t) \, \dd t & \textrm{for} \; k_{\epsilon}+k_{\beta} \; \textrm{odd} \,,
\end{array} \right.
\end{equation}
and $\mathcal{B}^{(2)}_{2,2}$ is given by
\begin{equation}
\mathcal{B}^{(2)}_{2,2} = x_o (-\tau) y_e (-\tau) \,.
\end{equation}
The set of probabilities~$\C_M$ with $M=0$, $1$, $2$, $3$, and~$4$ can be
chosen from among the sets in Table~\ref{table1} by choosing the frequency of
the analysis laser.  The angle $\theta$ can be varied continuously by changing
the inclination of the axis of linear polarization of the analysis laser.

In the expressions above the angle~$\theta$ is in the range $[0,\pi/2]$.  The
term in $\mathcal{B}_{2,2}^{(2)}$ is even in~$\theta$ about the midpoint
$\theta=\pi/4$, and the term in $\mathcal{B}^{(1,o)}_{2,2}$ and in the electron
EDM contribution~$D$ are both odd.  The odd linear combination for data for
different angles $\theta$,
\begin{equation}\label{plotsignal}
\begin{split}
\mathcal{P}^o_{\Delta} = \frac12 \biggl\{ &
\mathcal{P}^o \left(\frac{\pi}{16},
\frac{1}{\sqrt{2}} \,(\left| 4,4 \right> + \left| 4,-4 \right> ) \right)
\\
{}-{}&
\mathcal{P}^o \left(\frac{7\pi}{16},
\frac{1}{\sqrt{2}} (\left| 4,4 \right> + \left| 4,-4 \right> ) \right)
\biggr\}
\end{split}
\end{equation}
both cancels any contribution from $\mathcal{B}_{2,2}^{(2)}$ and maximizes the
sensitivity of what remains to $D$. The even linear combination isolates
$\mathcal{B}_{2,2}^{(2)}$. Once isolated,
$\mathcal{B}^{(1,o)}_{2,2}$ can be tuned to be of small magnitude because it
depends on the vertical component of the static magnetic field at the entrance
to the electric field plates, which can be tuned by varying a current in a
nearby coil.

Once $\mathcal{B}_{2,2}^{(2)}$ has been tuned to be of small magnitude,
canceled, or both, the remaining systematic $\mathcal{B}^{(1,o)}_{2,2}$ can be
canceled by taking the right linear combination of data for different initial
states, as shown by Eqs.~\eqref{Po44} and~\eqref{Po42}.   This systematic can
also be isolated, and then tuned to be of small magnitude by varying the
vertical component of the static magnetic field, in this case at a location
away from the entrance to the electric field plates.

A big advantage of the proposed observable is that there are no EDM mimicking
effects of
order~$1/\mu^3$, which had it been present would have been relevant on the
proposed level of sensitivity.  It is not yet necessary to cancel or control
EDM-mimicking effects of order~$1/\mu^4$, particularly since in practice, our
experimental accuracy will be limited by other systematic effects such as
magnetic Johnson noise, as for example discussed in Ref.~\cite{Mu2005}.

%
% EXPERIMENTAL REALIZATION
%
\section{EXPERIMENTAL REALIZATION}
\label{realism}

So far, we have considered a rather idealized model of an atomic
fountain. In order to make sure that an actual experiment
is not hindered by systematic effects arising from more realistic
conditions, we now consider systematic effects outside the model.

\begin{figure}[htb]
\begin{center}
\includegraphics[width=0.93\linewidth]{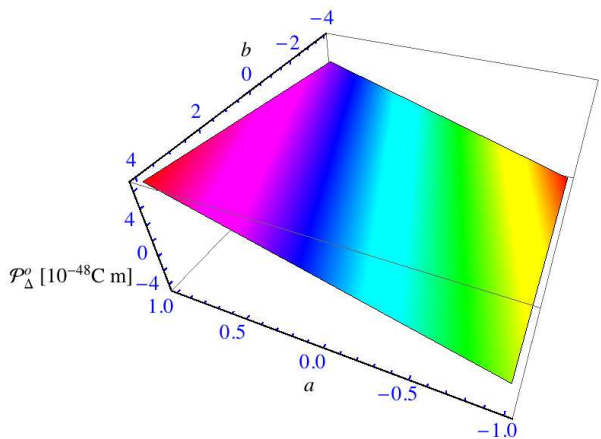}
\end{center}
\caption{\label{figg4} Plot of a numerical result for the
signal function given in Eq.~\eqref{plotsignal},
denoted as $\mathcal{P}^o_{\Delta}$, for ${}^{133}$Cs,
depending on $a$ and $b$. The $a$ and $b$ parameters describe
the magnetic field in the $y$ direction as
$\beta_y (t) = \beta_y (\mathrm{static}) \; ( a + b t^2)$.}
\end{figure}

The effect on $\mathcal{P}^o_{\Delta}$ of a linear vertical gradient in the
component of the magnetic field in the $y$-direction  is explored numerically
in Fig.~\ref{figg4}.   Because of the parabolic rise and fall of the atoms
while within the electric field, such a gradient results in the atomic rest
frame as a~$y$~component of the magnetic field that is of the form
$\beta_y({\rm static})(a+bt^2)$ for constants~$a$ and~$b$.  In
Fig.~\ref{figg4}, we use atomic data for the atom ${}^{133}{\rm Cs}$ and find
that~$\mathcal{P}^o_{\Delta}$ depends linearly on both~$a$ and~$b$.  The
numerical calculation which produces Fig.~\ref{figg4} has also been verified
against the analytic result found in Eq.~\eqref{B1o} for the EDM-mimicking
effect~$\mathcal{B}_{2,2}^{(1,o)}$.

\begin{figure}[htb]
\begin{center}
\includegraphics[width=0.93\linewidth]{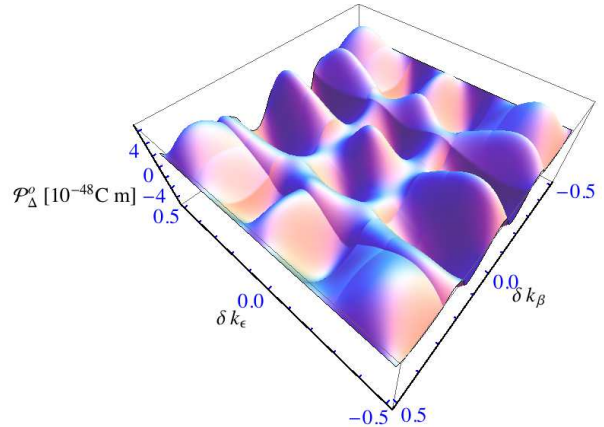}
\end{center}
\caption{\label{figg5} Plot of a numerical result for the signal of
Eq.~\eqref{plotsignal}, $\mathcal{P}^o_{\Delta}$, for ${}^{133}$Cs depending on
the deviations $\delta k_{\epsilon}$ of $k_{\epsilon}$ and $\delta k_{\beta}$
of $k_{\beta}$ from the phase adjustment in Eqs.~\eqref{muphaselock}
and~\eqref{bzphaselock}, respectively.}
\end{figure}

A similar numerical analysis is shown in Fig.~\ref{figg5}.of the dependence
of~$\mathcal{P}^o_{\Delta}$ on a deviation of~$\delta k_{\epsilon}$ and~$\delta
k_{\beta}$ from their respective phase adjustment to integer
multiples~$k_{\epsilon}$ and~$k_{\beta}$ of~$\pi$.  The results suggest that
the deviations should not exceed $\sim 10^{-3}$ for the error in the signal to
remain within the limits set forth by our target accuracy of~$\sim 10^{-9}$.

In a real as opposed to an ideal fountain, the common point of state
preparation and analysis does not occur at the edge of a step-function rise of
the electric field.  In practice both will occur at a point where the electric
field is essentially zero, well below the entrance to the electric field plates
where over a transition region whose vertical extent is the order of the plate
spacing the electric field ramps from zero to a constant value.  Between the
common point and the beginning of the transition region, the atoms will be
in the presence of a stray magnetic field that will effect a small rotation of
the atomic state.  In the transition region the combination of a significant
motional magnetic field and a stray magnetic field, acting while the tensor
Stark splittings are still small, generates among other effects an additional
rotation about the $z$ axis that can mimic the rotation effected by an electron
EDM.  All such effects influence the atom both on its way into the full
electric field and on its way out.

The additional effects can be studied by defining time-evolution operators for
each of these processes. For the transition region, we may expand the
time-evolution operator in terms of the short (dimensionless) transition
time~$\aleph$, which is small against the overall time~$T$ the atom spends in
the fountain. The rotation that occurs (about some arbitrary axis) before the
atom enters the transition region may be expanded in the small rotation
angle~$\delta$. We end up with three parameters $1/\mu$, $\aleph$,
and~$\delta$, which in the natural units of this work are all dimensionless,
small, and in a practical apparatus roughly of the same order of magnitude,
$\sim 1/\mu$. We can thus systematically model EDM-mimicking effects up to a
desired order in the small parameters.  Terms of third order are relevant on
the proposed level of sensitivity. A detailed discussion of this expansion,
which follows a formalism akin to the calculation of the time-evolution
operators considered here, remains beyond the scope of the current paper.
Technical details, which will be given elsewhere, reveal that the dominant
terms in~$\aleph$ and in~$\delta$ do not mimic an EDM and can thus be
discarded. The remaining effects can be satisfactorily described on the level
of our proposed target accuracy.

One might also speculate about the effect of a departure of the atomic motion
within the electric field from a straight line, as well as the effects of
misalignments of the axes of propagation or of polarization of the laser used
for state preparation and analysis. The first of these effects can be included
within our approach by adding an electric-field odd but time-even motional
magnetic field in the $y$ direction. We have carried out  a calculation of this
effect whose detailed discussion again is beyond the scope of the current
paper. We find that only the field-dependent functions change but not the
coefficient matrices. Our proposed cancellation mechanism is based solely on
the coefficient matrices and so not affected by a change in the values of the
field-dependent functions.  Misalignments of the direction of propagation of a
laser or of its axis of polarization can be modeled by representing the actual
laser as being an ideal laser rotated by a small angle; this small angle can be
treated as a small parameter that affects the time evolution of the atoms just
as do $1/\mu$, $\aleph$, and $\delta$.  We have not found any effects that
could possibly represent insurmountable difficulties at the level of our
proposed target accuracy; the design of the experimental apparatus is in
progress.

%
% FRANCIUM AND CESIUM SYSTEMS
%
\section{FRANCIUM AND CESIUM SYSTEMS}
\label{francium}

The availability of accelerator sources has made possible the use francium
(half-life 3 minutes) instead of cesium in the atomic fountain. The main
advantage of francium is that the enhancement factor~$R$ is about~$9$ times
larger in francium than in cesium (see Table~\ref{table3}). Furthermore,
certain francium isotopes, especially ${}^{211}$Fr, have a higher total angular
momentum, namely~$F=5$, which reduces the relative size of the EDM-mimicking
effects.  Moreover, the higher tensor polarizability also leads to~$\mu$ being
larger and thereby speeds the convergence of the formalism presented here.

Here, we present the results of our time-ordered calculation for~$F=3$, which
covers the case of~${}^{221}$Fr, and for~$F=5$,
which covers~${}^{211}$Fr. Only the
coefficient matrices $\mathcal{N}$ and $\mathcal{M}$ differ from those for
$F=4$; the corresponding field-dependent functions $\mathcal{B}$ and
$\mathcal{G}$ are identical.  In the asymptotic expansion of the unitary
transformation~$U$ to order~$1/\mu^2$,
those matrices not given for general~$F$
in Sec.~\ref{modelcesium} are gathered in the Appendices~E, F, G, and H
of Ref.~\cite{WuMuJe2012app}. Here,
we only present the results for our proposed
observable~$\mathcal{P}^o(\theta)[\Phi_0]$
for four initial states~$\big|\Phi_0\big>$
for the two isotopes; results for some other initial states
are given in Appendix~I of Ref.~\cite{WuMuJe2012app}.
\begin{widetext}
\begin{equation}\label{P33}
\begin{split}
\mathcal{P}^o &( \theta) \left[
\frac{1}{\sqrt{2}}\, \big(\big| 3,3 \big> + \big| 3,-3 \big> \big) \right] =
\sin (6 \theta) \biggl( 3 D - \frac{3 \mathcal{B}_{2,2}^{(1,o)}}{50 \mu^2} \biggr)
\frac{1}{8} \left( 15 \C_1 + \C_3 -10 \C_0 - 6 \C_2 \right) \\
& \quad\qquad\qquad\qquad\qquad\quad\;\, + \frac{15}{16}
\frac{\mathcal{B}_{2,2}^{(2)}}{\mu^2}
\biggl( 2 \sin (3 \theta) \cos ( \theta) \, \left[ \C_0 -\C_2 \right]
 + \sin (\theta) \cos (3 \theta) \,
\left[\C_3 -  \C_1 \right] \biggr) + \mathcal{O} (\mu^{-4}) \,,
\end{split}
\end{equation}
\begin{equation}\label{P55}
\begin{split}
\mathcal{P}^o& (\theta)
\left[\frac{1}{\sqrt{2}}\, \big(\big| 5,5 \big> + \big| 5,-5 \big> \big) \right] =
\sin (10 \theta) \biggl( 5 D - \frac{5 \mathcal{B}_{2,2}^{(1,o)}}{162 \mu^2} \biggr)\!
\frac{1}{128} \left( 210 \C_1 \!+ 45 \C_3 \!+ \C_5 \!-
126 \C_0 \!- 120 \C_2 \!- 10 \C_4 \right) \\
& \quad\quad + \frac{5}{7} \frac{1}{512} \frac{\mathcal{B}_{2,2}^{(2)}}{\mu^2}
\biggl( 2 \sin (5 \theta) \cos (3 \theta) \left[ 7 \C_0 -
4 \C_2 -3 \C_4 \right] + \sin (3 \theta) \cos (5 \theta)
\left[ 13 \C_3 + \C_5 -14 \C_1 \right] \biggr) + \mathcal{O} (\mu^{-4}) \,,
\end{split}
\end{equation}
\begin{equation}\label{P30}
\begin{split}
\mathcal{P}^o (\theta)
\left[ \big|3, 0 \big> \right] &=
\sin (2 \theta) \frac{\mathcal{B}_{2,2}^{(2)}}{\mu^2}
\frac{5}{4} \left( \C_3 - \C_1 \right) + \mathcal{O} (\mu^{-4}) \,,
\end{split}
\end{equation}
\begin{equation}\label{P50}
\begin{split}
\mathcal{P}^o (\theta)
\left[ \big|5, 0\big>  \right]  &=
\sin (2 \theta) \frac{\mathcal{B}_{2,2}^{(2)}}{\mu^2}
\frac{35}{16} \left( 3 \C_5 -\C_3 -2 \C_1 \right) + \mathcal{O} (\mu^{-4}) \,,
\end{split}
\end{equation}
\end{widetext}
The variable~$D$  has been defined in Eq.~\eqref{defD}
and incorporates the effect of the  EDM in view of its dependence
on $\sigma_F$ defined in Eq.~\eqref{sigmaF}.
Equations~\eqref{P33} and~\eqref{P55} show that when~$F$ is odd, initial states
consisting of superpositions of states with equal $|M|$ do not
exhibit the symmetry in~$\theta$ around~$\pi/2$ that when~$F$ is even could be
used to cancel the systematic effect~$\mathcal{B}_{2,2}^{(2)}$. Measurements on
the states $\big|F, 0\big>$ are shown by equations~\eqref{P30} and~\eqref{P50} to
be sensitive only to only to $\mathcal{B}_{2,2}^{(2)}$, which can therefore be
measured and tuned to be small by changing the vertical component of the static
magnetic field at the entrance to the electric field plates. Any residual
effect can be separated from the values of $D$ and of the other
systematic~$\mathcal{B}_{2,2}^{(1,o)}$ either by combining data for different
angles $\theta$, or by altering the laser frequency and combining data for
different sets of values~$\C_M$.  The sets accessible for $F=3$ and for $F=5$
may be found in~Appendix~B of Ref.~\cite{WuMuJe2012app}.

%
% CONCLUSIONS
%
\section{CONCLUSIONS}
\label{conclusions}

We set up the formalism for the theoretical description of the quantum dynamics
of an alkali atom within an atomic fountain designed for an EDM experiment.
The low velocity of the atoms inside a fountain reduces the motional magnetic
field, which arises as the Lorentz transformation of the applied electric
field, by a factor of~$100$ compared to experiments on thermal atomic beams.
In an atomic fountain the quantization axis is defined by the direction of the
externally applied strong electric field, and is not defined by a magnetic
field; we are therefore free to greatly reduce all magnetic fields and their
attendant systematic errors by use of extensive magnetic shielding and nulling
coils.  Compared to many previous experiments these two features suppress
effects that mimic the presence of an EDM, because such effects scale linearly
with both the motional magnetic field and any stray field.  We describe a
theoretical calculation that identifies the remaining EDM-mimicking effects and
devise schemes to eliminate them.

A crucial part of our formalism is writing the time evolution operator from
time-ordered perturbation theory in terms of an analytic expansion in the
inverse number of electric-field induced Rabi oscillations within the hyperfine
manifold.  When the magnetic fields in the~$x$ and~$y$ directions are zero, the
time-evolution operator for a single hyperfine manifold is a diagonal matrix of
phases $V$, which reduces to a simple rotation of the atomic system provided
the cumulative electric and magnetic field phases of Eq.~\eqref{muphaselock}
and Eq.~\eqref{bzphaselock} are tuned to integer multiples of~$\pi$.

The perturbing effects of stray magnetic fields in the~$x$ and~$y$~directions
as well as the motional magnetic field in the~$x$~direction, still without an
EDM term, can be treated when the electric field is constant by expanding the
correction term $W(t)$ in the formula for the time-evolution operator $U(t) =
V(t) \, W(t)$ in an asymptotic expansion in inverse powers of $\mu =
\epsilon_z^2$, where $\mu/(2\pi)$ describes the number of Rabi oscillations in
the hyperfine manifold induced in the absence of perturbing fields. Under
typical experimental conditions a value of~$\mu$ on the order of one hundred
can be obtained and so expansion is therefore rapidly converging. The expansion
of $W$ to order $1/\mu^2$ requires consideration of terms of fourth order of
perturbation theory; we present the expansion in terms of constant matrices
multiplied by analytic functions of the stray and motional magnetic fields.

We then define an observable $P(\theta)$, which is a function of the angle of
inclination of the linear polarization of the laser used to analyze the final
state of the atom.  This observable is sensitive to a rotation of the initial
state about the electric field axis, as would be induced by the presence of an
electron EDM.  A transformation of the final state into pieces symmetric and
antisymmetric with respect to two transformations~$B$ and~$S$ defined in
Eqs.~\eqref{Bdef} and~\eqref{Sdef} proves that many contributions cancel.
Critically it is proven that in the asymptotic expansion of $P$ only even
powers of $1/\mu$ appear, so knowledge of the time development operator out to
order $1/\mu^2$ suffices to compute $P$ with error $1/\mu^4$, not $1/\mu^3$.

After taking the difference for opposite signs of the electric field, besides
the effect of an electron EDM only two systematic errors survive, which can
both be isolated and canceled by combining data for polarization angles, laser
frequencies, and initial states; moreover both systematic errors once measured
can be tuned to be of small magnitude by imposing additional small magnetic
fields.  The systematics start intrinsically smaller than in previous
experiments because of the smaller velocity of atoms in an atomic fountain,
which reduces the magnitude of the motional magnetic field, and because in an
atomic fountain with a large electric field a magnetic field is not required to
define a quantization axis, so stray components of the magnetic field can be
suppressed to the limit provided by the surrounding magnetic shielding.

The proposed apparatus is expected to detect an EDM on a level of $2 \times
10^{-50}\,{\rm C\,m}$, or better. The limit is due to higher-order effects and
due magnetic Johnson noise~\cite{Mu2005} from the materials used in the
apparatus.

%
% Acknowledgments
%
\section*{Acknowledgments}

The authors thank H.~Gould and B.~Feinberg for insightful conversations and
B.J.W.~thanks the Lawrence Berkeley Laboratory for warm hospitality during a
visit in early 2011.  U.D.J.~and B.J.W.~acknowledge support from the National
Science Foundation (Grant PHY-1068547) and by a precision measurement grant
from the National Institute of Standards and Technology.  This work was also
supported by the Director, Office of Science, of the U.S.~Department of Energy
under Contract No. DE-AC02-05CH11231.

\end{document}